%% file: 0.TPDS.IEEE.main.tex
\documentclass[lettersize, journal]{IEEEtran}
\input{0.package}
\input{0.macro}
\usepackage{cite}

\begin{document}

\bstctlcite{bstctl:nodash}
\title{\fontsize{22pt}{24pt}\selectfont{
\input{tex/0.title}

}
}

\input{tex/0.authors}



\maketitle

\begin{abstract}
\input{tex/0.abstract}
\end{abstract}


\input{tex/1.body}

\bibliographystyle{IEEEtran}
\bibliography{ref}

\input{tex/2.bio}

\end{document}

%% file: 0.package.tex
\usepackage{listings}
\lstdefinestyle{promptstyle}{
    basicstyle=\ttfamily\small,
    frame=single,
    rulecolor=\color{black},
    breaklines=true,
    numbers=none,
    tabsize=4,
    keywordstyle=\color{blue},
    commentstyle=\color{green!50!black},
    stringstyle=\color{red}
}

\usepackage{longtable}
\usepackage{booktabs}   

\usepackage{svg}
\usepackage{booktabs}
\usepackage{tabularx}
\usepackage{tikz}
\usepackage{comment}

\usepackage{array}
\usepackage{mathtools}
\usepackage{multirow}
\usepackage{multicol}
\usepackage{hhline}
\usepackage[export]{adjustbox}
\usepackage{graphicx}
\usepackage{makecell}

\usepackage{threeparttable}
\usepackage{url}

\usepackage[ruled,vlined]{algorithm2e}
\usepackage{algpseudocode}
\usepackage[skip=2pt]{subfig}
\usepackage{todonotes}
\usepackage{enumitem}

\usepackage{tikz}
\usetikzlibrary{calc}

\usepackage{hhline}
\usepackage{wrapfig}
\usepackage{nicematrix} 
\usepackage[table,xcdraw]{xcolor}

\usepackage{bm}
\usepackage{xr-hyper}
\usepackage{pifont}
\newcommand{\xmark}{\ding{55}}%

%% file: 0.macro.tex
\newcommand{\floor}[1]{\left\lfloor #1 \right\rfloor}

\newcolumntype{|}{!{\vrule width 0.5pt}}
\newcolumntype{?}{!{\vrule width 1pt}}
\newcolumntype{^}{!{\vrule width 1.2pt}}

\definecolor{myblue}{RGB}{91,155,213}
\definecolor{mydark}{RGB}{0,0,0}

\newcommand{\emh}[1]{\textit{#1}}

\SetKwInOut{Input}{Input}
\SetKwInOut{Output}{Output}
\SetKwInOut{Data}{Data}
\SetKwProg{Tree}{Tree}{}{EndTree}

\algtext*{EndWhile}
\algtext*{EndIf}

\newcommand{\Xuetao}[1]{\textcolor{black}{#1}}

\newcommand{\hC}{\rowcolor[HTML]{333333}}
\newcommand{\tH}[1]{\multicolumn{1}{c}{\textcolor{white}{#1}}}

\newcommand{\TSS}[1]{\textsuperscript{#1}}
\newcommand{\method}{\texttt{edgeFBP}}

%% file: tex/0.title.tex


An Efficient Out-of-Core Tomographic Imaging Framework for Edge Devices

%% file: tex/0.authors.tex
\author{
  \IEEEauthorblockN{
    Xuetao Chen$^{*}$,
    Cong Ma$^{*}$,
    Xiangyu Meng$^{\dagger}$,
    Du Wu$^{\dagger}$,
    Zhengyang Bai,
    Tao Luo, 
    Zhaorui Zhang,
    Emmanuel Jeannot,
    Edgar Josafat Martinez Noriega,
    Xun Wang,
    Peng Chen,
    Amelie Chi Zhou,
    Mohamed Wahib
  }
  
  
\thanks{$^{*}$Co-first authors. $^{\dagger}$Co-corresponding authors.}
\thanks{This work was supported by the National Key Research and Development Program (2025YFB4507000), the National Natural Science Foundation of China (Grant Nos. 61972416, 62272479), Taishan Scholarship (tstp20240506, tsqn202408087), Natural Science Foundation of Shandong Province (ZR2022LZH009), National Research Foundation, Singapore (NRF), and the Ministry of Digital Development and Information (MDDI) under the AI Visiting Professorship (Award No. AIVP-2025-005).}
\thanks{Xuetao Chen and Amelie Chi Zhou are with Hong Kong Baptist University, Kowloon, Hong Kong, China (email:\{csxtchen, amelieczhou\}@comp.hkbu.edu.hk).}
\thanks{
Cong Ma is with the Hokkaido University, Sapporo, Japan, 060-0814. (email:cong.ma.m1@elms.hokudai.ac.jp).
}
\thanks{
Xiangyu Meng and Xun Wang are with the College of Computer Science and Technology, China University of Petroleum (East China), Qingdao, China, 266580 (email:x\_meng0420@163.com, wangsyun@upc.edu.cn).
}
\thanks{
Du Wu, Zhengyang Bai, Peng Chen, and Mohamed Wahib are with the RIKEN Center for Computational Science, Kobe, Japan, 650-0047 (email:\{du.wu, zhengyang.bai, peng.chen, mohamed.attia\}@riken.jp)
}
\thanks{
Tao Luo is with the A*STAR Institute of Advanced Intelligence and Computing (A*STAR IAIC), 
Singapore, 138632 (email:luo\_tao@a-star.edu.sg).
}
\thanks{
Zhaorui Zhang is with the Hong Kong Polytechnic University, Hung Hom, Kowloon, Hong Kong, China (email:zhaorui.zhang@polyu.edu.hk)
}
\thanks{
Edgar Josafat Martinez Noriega is with the National Institute of Advanced Industrial Science and Technology (AIST), Tsukuba, Japan, 305-8560 (email:edgar.martineznoriega@aist.go.jp)
}
\thanks{
Emmanuel Jeannot is with the Inria, Univ. Bordeaux, LaBRI,Talence, France, 33405 (email:emmanuel.jeannot@inria.fr)
}
}


%% file: tex/0.abstract.tex
Computed Tomography (CT) is an essential 3D imaging technology widely used in medical diagnostics and scientific research. However, performing CT imaging on edge devices is challenging due to limitations in computational power, memory capacity, and energy budget.
This paper presents an efficient CT reconstruction framework, called \method{}, designed for Nvidia Jetson System-on-Chip (SoC) devices.
\method{} adopts an end-to-end pipeline design for efficient out-of-core image reconstruction under tight power and memory constraints. 
\method{} utilizes a mixed-precision strategy leveraging half-precision Tensor Cores(TCs) to accelerate the bottleneck back-projection(BP) kernel.
\method{} achieves a $1.83\times$ speedup over the widely used RTK library on Jetson Nano and a $2.56\times$ speedup on Jetson AGX.
Under a strict 25-Watt power budget, \method{} on Jetson Nano achieves up to 5$\sim$48$\times$ higher energy efficiency than an Nvidia DGX A100, enabling datacenter‑scale imaging on constrained edge devices.

\begin{IEEEkeywords}

\input{tex/0.keywords}

\end{IEEEkeywords}




%% file: tex/0.keywords.tex
Computed Tomography, image reconstruction, Nvidia Jetson, GPU, System-on-Chip.

%% file: tex/1.body.tex
\section{Introduction}

Computed Tomography (CT)~\cite{kak2001principles,natterer2001mathematics,kalender2011computed,feldkamp1984practical} is a vital 3D imaging technique widely used in medical diagnostics, industrial inspection, and scientific research, targeting different objects and applications. By acquiring multiple X-ray projections from different angles, the Filtered Back-Projection (FBP) image reconstruction algorithm generates cross-sectional images of scanned objects. High-quality CT reconstruction requires intensive computations involving a large number of matrix operations and memory-intensive processing of projection data. Traditionally, these workloads have been handled by various types of accelerators, including ASICs~\cite{wu1991asic}, multi-core CPUs~\cite{Serrano2014Heterogeneous}, SIMD-optimized DSPs~\cite{treibig2013pushing, hofmann2014comparing}, high-end GPUs~\cite{Chen2019iFDK,hidayetoglu2020petascale,8425161}, and distributed systems~\cite{Wang2017Massively3D,Wang2022Ptycho,Wang2021Physics}. However, the growing need for real-time and portable imaging solutions has created a demand for efficient CT reconstruction on edge devices, which operate under strict constraints in computational power, memory capacity, and energy efficiency.


There is a strong motivation to optimize the FBP algorithm for System-on-Chip (SoC) edge devices, such as Nvidia Jetson series, to leverage their power efficiency and computational capability, as summarized in Table~\ref{tbl:jetson}.
While high-end discrete GPUs have long been the standard for image reconstruction~\cite{Palenstijn2015Distributed,lu2016cache}, their substantial power and space requirements are unsuitable for modern portable and embedded CT systems. Recent trends toward compact and mobile micro-CT devices~\cite{clark2014micro,compactCT,ying2017micro} demand solutions that are both energy-efficient and cost-effective.
For instance, typical micro-CT systems operate within a 450-watt power budget~\cite{xth450,bruker_skyscan1276}, making it impractical to integrate power-hungry GPUs, such as Nvidia’s A100 or H100. In contrast, Jetson modules offer integrated GPU acceleration with significantly lower power consumption and smaller form factors, enabling efficient on-device reconstruction. Specifically, these SoC edge devices simplify system integration by consolidating compute, memory, and I/O into a single compact platform. 
These features make Jetson platforms attractive for deploying image reconstruction applications in real-world CT systems.

\input{tables/JetsonDevicesTable}
Image reconstruction on resource-constrained edge devices, such as Jetson modules, faces several key challenges. The commonly used FBP algorithm involves two compute-intensive kernels: filtering and back-projection (BP), which impose high demands on computation and memory resources, and are inherently limited on Jetson platforms in Table~\ref{tbl:jetson}. Optimizing these kernels requires effective use of the Jetson heterogeneous architecture, including CPUs, GPUs, and specialized matrix engines (namely Tensor Cores), which requires algorithmic adaptations to fully exploit this architecture.
Moreover, high-resolution reconstruction (i.e. $2048^3$) drastically increases memory usage, exceeding the available on-device memory. It is essential to design a computing pipeline that supports efficient out-of-core processing.



We propose an efficient FBP framework, called \method{}, for out-of-core image reconstruction on Jetson.
\method{} leverages the heterogeneous computing capabilities of the CUDA platform to optimize FBP imaging computations. We utilize the highly optimized \emph{cuFFT} vendor library to accelerate the filtering computation, which is critical to overall FBP throughput.
We design a novel BP algorithm that reformulates the computationally intensive coordinate projection computations into matrix operations suitable for execution on Tensor Cores (TCs). Specifically, we perform matrix multiplications using FP16 arithmetic while recovering single-precision accuracy through mixed-precision computation.
As shown in Tab~\ref{tbl:jetson}, TCs acceleration enables power-constrained edge devices to deliver significant performance improvement. Since the compute-intensive BP stage is often the primary performance bottleneck in FBP, TCs are essential for optimizing BP and pushing the performance limits of edgeFBP on Jetson devices.
To support high-resolution 3D image reconstruction, we adopt an out-of-core approach by splitting projections (input) and volumes (output) into smaller blocks to fit the limited memory capacity of Jetson modules, and design a pipeline scheduling strategy to overlap I/O operations with FBP computation.


Evaluations on Jetson modules demonstrate the power and computational efficiency of \method{} with the out-of-core image reconstruction capability.
Roofline and throughput comparisons show that \method{} on Jetson Orin Nano outperforms RTK by an average of 1.83$\times$ and approaches the computational limits of the device, while achieving a 2.56$\times$ throughput improvement on Jetson AGX Orin.
In terms of energy efficiency, under a strict 25 W power budget, edgeFBP achieves up to 5$\sim$48$\times$ higher energy efficiency than an Nvidia DGX system (8$\times$A100 server), highlighting the effectiveness of hardware-aware optimization for CT reconstruction on power-constrained edge devices.

The contributions of this paper are summarized as follows:
{
\setlength{\leftmargini}{15 pt}
    \begin{itemize}
        \item This work is the first to perform CT image reconstruction on the Jetson series, a SoC platform typically used for robotics and machine learning. This work paves the way \Xuetao{for} enabling an unprecedented data center-grade $2048^3$+ resolution CT on micro-CT devices, within minutes.
        \item We propose an end-to-end image reconstruction framework for Jetson platforms, leveraging their heterogeneous architecture to enable efficient out-of-core reconstruction within the memory constraints of edge devices.
        \item We develop a novel back-projection algorithm that leverages TCs with FP16 mixed-precision computation, adopted for efficient coordinate computation while maintaining image quality requirements.
        \item We present a performance evaluation to demonstrate both computational and power efficiency, as well as the capability for out-of-core image reconstruction.
    \end{itemize}
}



\section{Background \& Related Work}\label{sec:background}
This section introduces the background of Cone-beam Computed Tomography (CBCT), the Filtered Backprojection (FBP) reconstruction algorithm, and related work.

\subsection{Cone-beam Computed Tomography (CBCT)}
\begin{figure}[t]
    \centering
    \includegraphics[width=\columnwidth]{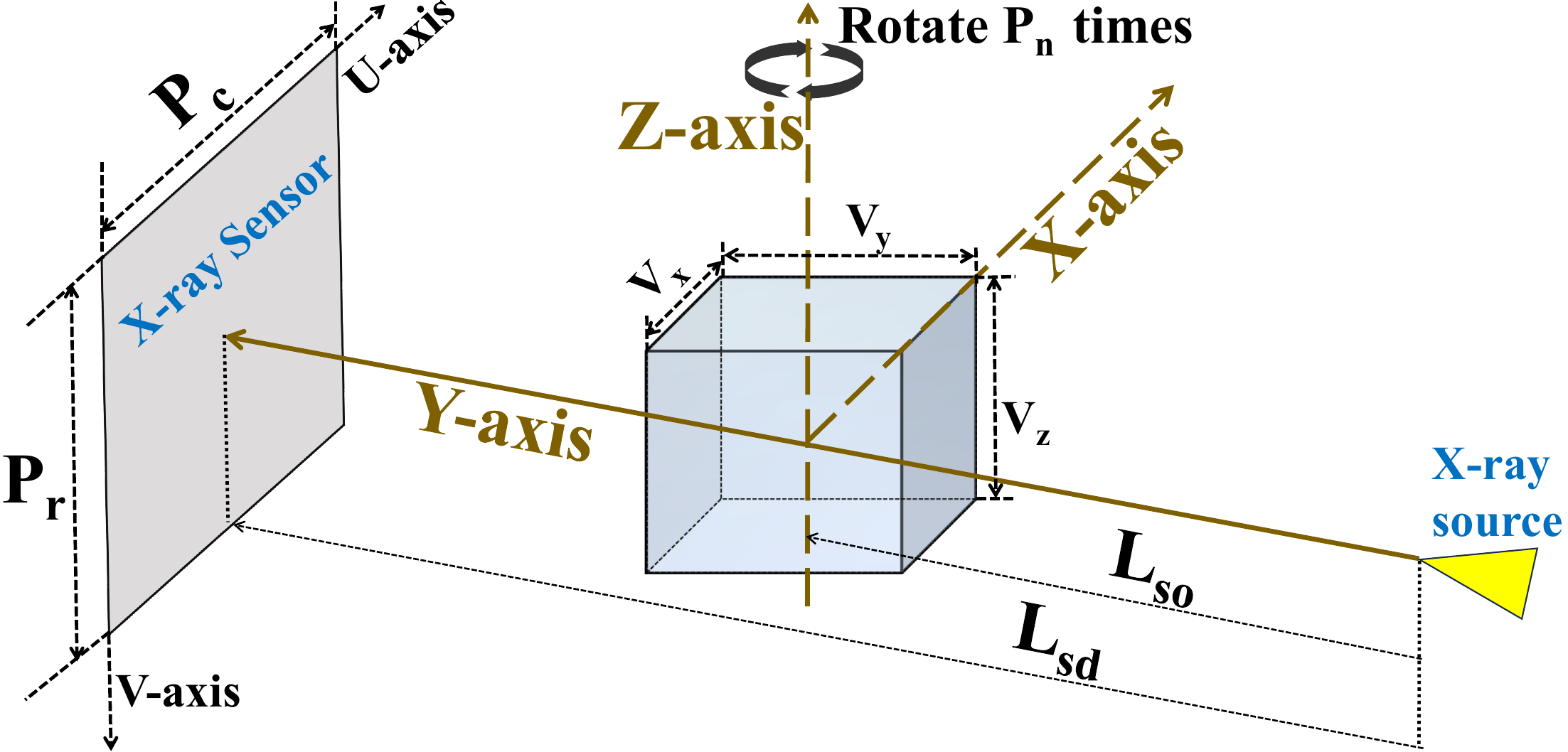} 
    \caption{Geometry of the Cone-beam Computed Tomography (CBCT) imaging system. CBCT consists of a micro-focus X-ray source and a flat-panel detector (or X-ray sensor) with $P_r$ rows and $P_c$ columns of sensing elements.}
    \label{fig:cbct}
\end{figure}

As illustrated in Fig.~\ref{fig:cbct}, the system includes a micro-focus X-ray tube as the source and a flat-panel detector (or X-ray sensor). The distance from the X-ray source to the rotation axis (Z-axis) is denoted as ${L_{so}}$, and the distance from the source to the detector is ${L_{sd}}$.
The detector contains $P_r$ rows and $P_c$ columns of sensing elements (pixels). Its U-axis is aligned with the system’s X-axis, and its V-axis is aligned with the Z-axis. The reconstructed 3D volume is defined by $V_x$, $V_y$, and $V_z$ voxels along the X, Y, and Z directions, respectively. These volume elements, referred to as voxels, denote the reconstructed 3D image.
The $3{\times}4$ projection matrix $M_\psi$ is formed by multiplying the following matrices:
\begin{equation*}
\begingroup
\setlength{\arraycolsep}{1pt} 
\renewcommand{\arraystretch}{1.1} 
\resizebox{\columnwidth}{!}{$
M_{\psi}=
\begin{bmatrix}
    \dfrac{L_{sd}}{\lambda_c} & 0 &
    \dfrac{P_c-1}{2}+\omega_c & 0\\
    0 & \dfrac{L_{sd}}{\lambda_r} &
    \dfrac{P_r-1}{2}+\omega_r & 0\\
    0 & 0 & 0 & L_{sd}
\end{bmatrix}
\begin{bmatrix}
    \cos(\psi) & -\sin(\psi) & 0 & \omega_{cor}\\
    0 & 0 & -1 & 0\\
    \sin(\psi) & \cos(\psi) & 0 & L_{so}\\
    0 & 0 & 0 & 1
\end{bmatrix}
\begin{bmatrix}
    \lambda_x & 0 & 0 &
    \dfrac{1-\lambda_x V_x}{2}\\
    0 & \lambda_y & 0 &
    \dfrac{\lambda_y V_y}{2}\\
    0 & 0 & \lambda_z &
    \dfrac{\lambda_z V_z}{2}\\
    0 & 0 & 0 & 1
\end{bmatrix}
$}
\endgroup
\end{equation*}
$M_\psi$ is essentially used to map the spatial position of each voxel onto the 2D plane of the X-ray sensor.

\input{tables/Parameters}

\subsection{FBP Image Reconstruction for CBCT}\label{sec:fbp}
This section briefly presents the 3D image reconstruction for CBCT by the FBP algorithm as presented in~\cite{Jaffray2000FlatPanel}.

\paragraph{Filtering}
The filtering applies a one-dimensional Ramp filter~\cite{Kak1988Principles} to each row of pixels in the 2D projection. The filtering computation can be written as:
\begin{equation}\centering
\label{eqn:flt}
\small
Q(u, v) = \left( {L_{sd}}/{\sqrt{D(u, v)^2 + L_{sd}^2}} \cdot P(u, v) \right) \ast R(u)
\end{equation}
where $\tiny{D(u, v)^2 = \left( \Delta u \left( u - {P_c}/{2} \right) \right)^2 + \left( \Delta v \left( v - {P_r}/{2} \right) \right)^2}$, $\ast$ denotes convolution, 
and $R(u)$ indicates the ramp filter~\cite{zeng2014revisit}. To perform the filtering in the frequency domain, a convolution is performed using the Fast Fourier Transform (FFT)~\cite{Brigham1988FFT}, which significantly reduces the filtering complexity.

\paragraph{Back Projection}
Alg.~\ref{alg:BP} illustrates the 3D cone-shaped back-projection implementation used in the RTK library~\cite{rit2014reconstruction}.
The input includes the filtered projection images \emph{Q}, with dimensions $P_n{\times}P_r{\times}P_c$, and the corresponding projection matrices \emph{M}, of size $P_n{\times}4{\times}3$. Each projection matrix is defined as $M[s] = M_{\psi}$, where $s{\in}[0, P_n)$ and the projection angle is calculated as $\psi = 360 \cdot t / P_n$ for a full circular scan.
The back-projection process follows the standard projection model: for each voxel at position $[i, j, k]$, the 3D coordinate is projected into 2D detector space using the matrix-vector multiplication $M_{\psi} \cdot [i, j, k, 1]^T$, yielding projected coordinates $[x, y, z]^T$ (Alg.~\ref{alg:BP}, line 6). The $x$ and $y$ coordinates are then normalized by dividing by $z$.
The projection image is sampled at sub-pixel coordinates $(x, y)$ using bilinear interpolation, implemented by the \emph{interp2} function (lines 9–14). The interpolated value is weighted by $1/z^2$ and accumulated into the output volume at $I[k][j][i]$, with the $z$ value accounting for geometric scaling in the cone-beam geometry.
The \emph{interp2} function performs bilinear interpolation by computing weighted sums of four neighboring pixel values~\cite{russ1990image}, based on the sub-pixel offsets $(\varepsilon_u, \varepsilon_v)$.
As the algorithm iterates over all projections and all voxels in BP, its computational complexity is $\mathcal{O}(N^4)$.

\begin{figure*}[t]
    \centering
    \includegraphics[width=\linewidth]{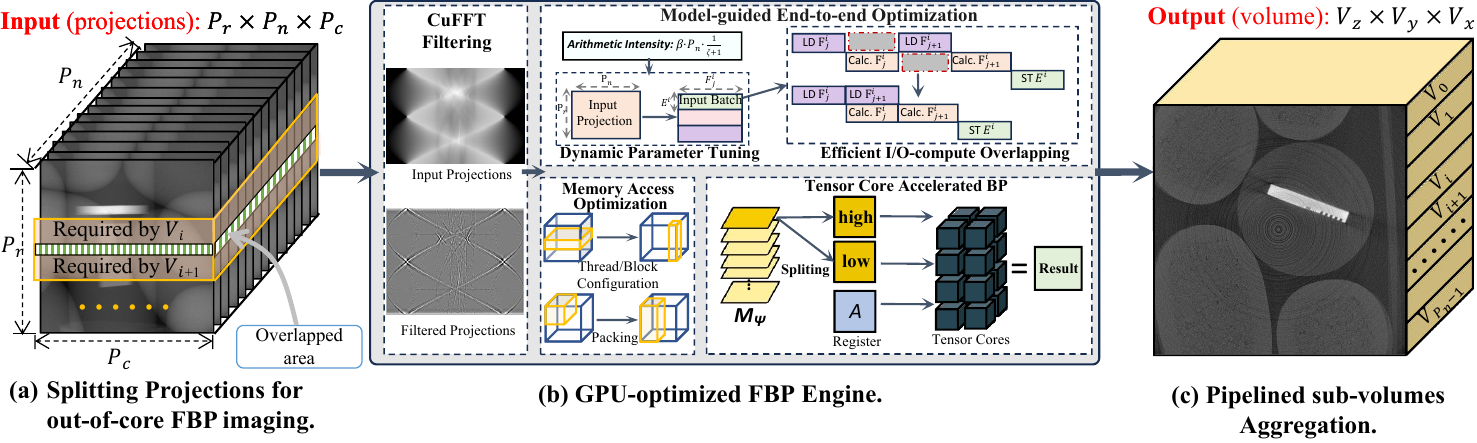} 
    \caption{
    Overview of \method{}. As a highly optimized end-to-end pipeline for out-of-core CT imaging on edge devices, the workflow \textbf{(a)} partitions the projections into smaller batches for out-of-core processing, and \textbf{(b)} reconstructs each batch independently using an optimized FBP engine with GPU memory access optimization and TCs-accelerated computation. Then the framework \textbf{(c)} aggregates the sub-volumes (e.g., $V_0$, $V_1$, etc.) to construct the final 3D volume in a pipelined fashion.
    }
    \label{fig:overview}
\end{figure*}
\subsection{Related work}

\input{Algorithms/backprojection}

CT image reconstruction is computationally intensive, with existing approaches primarily optimizing computations on discrete accelerators, clusters, and supercomputers rather than edge devices.
The Reconstruction Toolkit (RTK)~\cite{rit2014reconstruction}, provides standard implementations of FBP on both CPU and GPU platforms. Extending RTK to distributed architectures has also been investigated. Palenstijn et al.~\cite{Palenstijn2015Distributed} developed a distributed SIRT implementation for GPU clusters, while Cui et al.~\cite{Cui2013Distributed} proposed DMLEM for multi-GPU systems. 
Bicer et al.~\cite{hidayetouglu2021memxct,hidayetoglu2020petascale,hidayetouglu2019memxct} and Xiao et al.~\cite{Wang2017Massively3D,wang2019consensus} scale large-scale CT reconstruction with iterative solvers on supercomputers.
Meanwhile, frameworks like TIGRE~\cite{biguri2019arbitrarily} offer efficient GPU-based implementations of iterative methods, such as MLEM~\cite{Shepp1982ML} and SIRT~\cite{gregor2008computational}.
However, these solutions are often limited to small-volume reconstructions due to constraints in hardware memory capacity.
While some systems~\cite{lu2016cache, QUINTANAORTI2022106725} employ cache-aware optimizations for out-of-core reconstruction, they do not provide advanced Tensor Core support or fine-grained memory management for SoC GPU devices.

Jetson platforms are widely used in applications that require high performance within strict power and memory constraints. Their small form factor integrated CPU and GPU make them well-suited for various tasks such as Deep Learning-based object recognition~\cite{Dlrecon}, SLAM \cite{slam}, and medical image processing \cite{SWAMINATHAN2025906}. These emerging architectures require dedicated designs to optimize memory usage and power efficiency. Matsubara et al. \cite{9842809} reduced power by 49\% and latency by 89\% using BottleFit. Bouwmeester et al.~\cite{10161258} enabled real-time optical flow for nano-drones using NanoFlowNet. Bakhtiarnia et al. \cite{10096914} demonstrated dynamic split computing for variable bandwidth. Han et al. further extend the capability of Jetson platforms toward large-model workloads like AWQ, which enables activation-aware quantization for efficient on-device LLM inference in extremely resource-constrained environments~\cite{MLSYS2024_AWQ}. These studies demonstrate Jetson can handle real-time processing in embedded and portable systems with limited computing resources.

Despite these advances, the majority of prior work targets high-end computing platforms, which are not suitable for modern embedded or portable CT systems due to power and memory constraints. As CT applications increasingly demand high-resolution outputs within compact, low-power systems, memory capacity becomes a major bottleneck. To address this challenge, volume decomposition and out-of-core strategies have been proposed~\cite{lu2016cache,biguri2019arbitrarily}, but few frameworks support high-resolution reconstruction on edge devices.

\method{} introduces a full-stack design and optimization for Nvidia Jetson modules. 
This work is the first to perform CT image reconstruction on the Jetson series, a SoC platform typically used for robotics and machine learning, achieving efficient and scalable reconstruction, enabling high-resolution reconstruction for portable CT devices.

\section{Proposed Imaging Framework: \method{}} \label{sec:framework}
\textbf{Overview of the proposed \method{}}. We propose \method{}, an efficient out-of-core image reconstruction framework designed for resource-constrained edge devices. Fig.~\ref{fig:overview} presents an overview of \method{}. \method{} co-designs data management, execution scheduling, and GPU kernels to address both memory and compute limitations. Specifically, it partitions projection and volume data to enable out-of-core image reconstruction, employs a throughput-oriented pipeline that overlaps storage I/O with computation, optimizes GPU memory-access patterns to improve data locality, and leverages TCs to accelerate the BP kernel, substantially reducing end-to-end imaging time.

\subsection{Out-of-core Projection \& Volume Partition Design} \label{sec:Out-of-Core}
\begin{figure}[t]
    \centering

    \begin{minipage}{0.665\columnwidth}
        \centering
        \subfloat[$E$ splitting.]{
            \includegraphics[width=\linewidth]{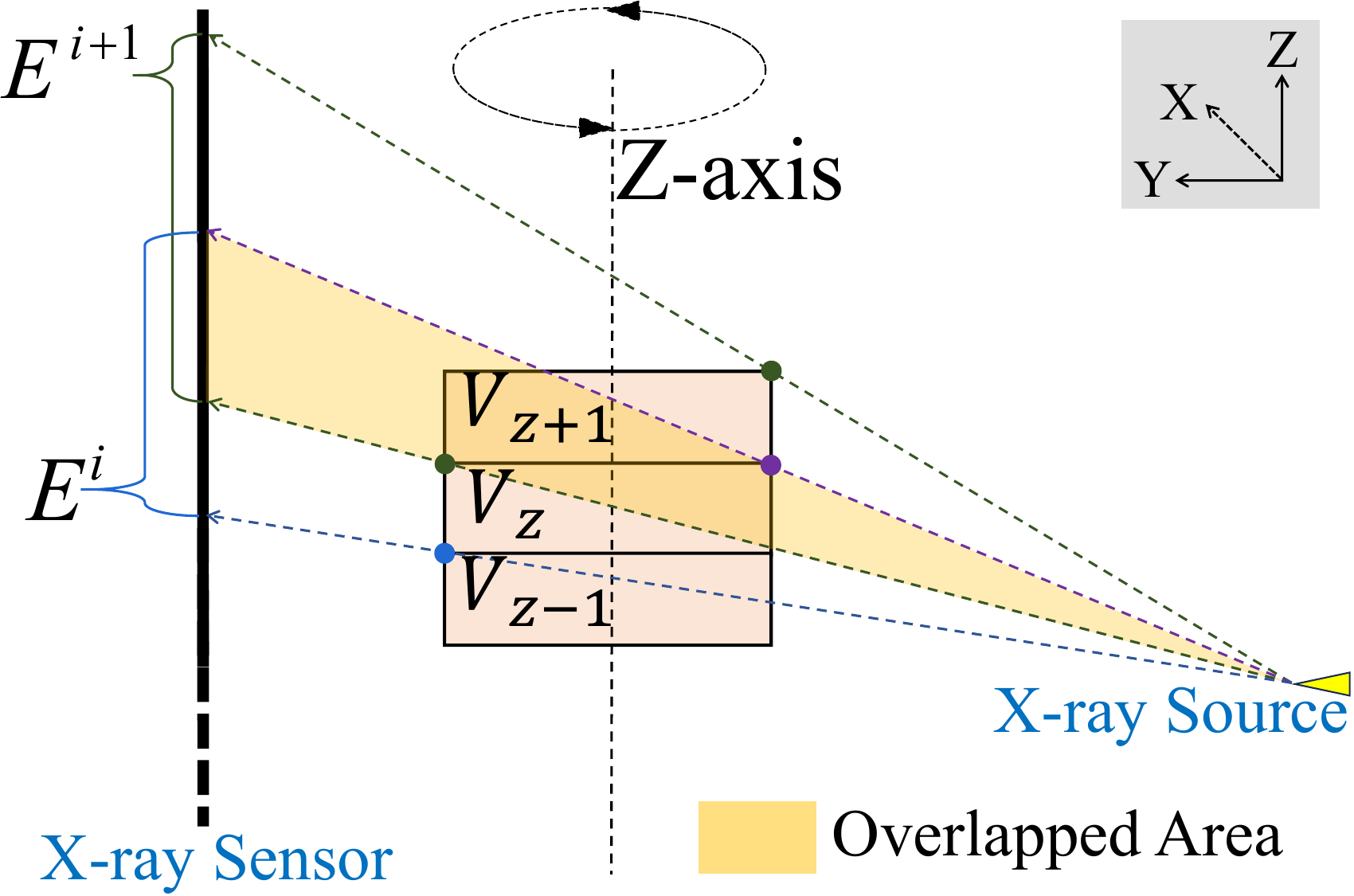}
            \label{fig:overlappeda}
        }
    \end{minipage}
    \hfill
    \begin{minipage}{0.311\columnwidth}
        \centering
        \subfloat[$F$ splitting.]{
            \includegraphics[width=\linewidth]{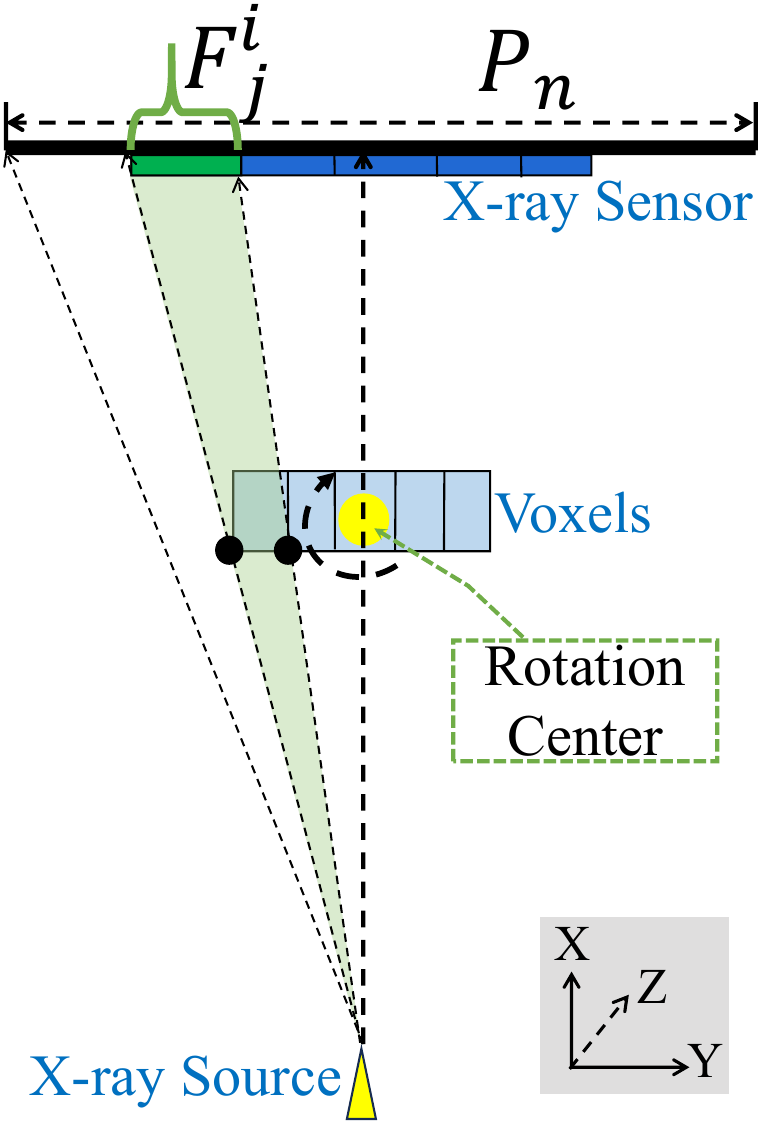}
            \label{fig:overlappedb}
        }
    \end{minipage}
    
    \begin{tikzpicture}[remember picture, overlay]
        \draw[dashed, thick] ($(current page.center) + (6.35cm, -0.2cm)$) -- ++(0,4.2cm);
    \end{tikzpicture}

    \caption{Our projection and volume partitioning design for out-of-core FBP reconstruction in \method{}. A dynamic two-level batching strategy is introduced. \textbf{(a)} Outer batch $E$ is partitioned along $V_z$. \textbf{(b)} Inner batch $F$ is dynamically partitioned along $P_n$ according to the available memory capacity.} 
    \label{fig:overlap}
\end{figure}
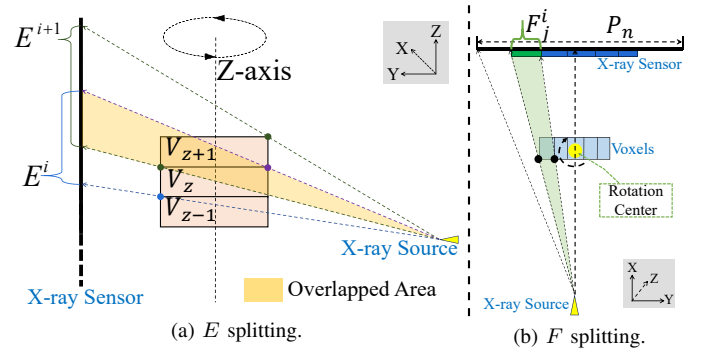

Due to limited memory on edge devices, it is infeasible to load all projection data and compute the entire output volume at once, making out-of-core processing essential. 
As shown in Fig.~\ref{fig:overlap}, we introduce a dynamic two-level batching out-of-core computational strategy. 
In this optimization strategy, the outer batching operates along the axial depth ($V_z$), guided by a balance between I/O and computation. This approach naturally matches the slice-wise structure of the reconstructed volume and reduces the memory footprint of both the input projections and the volume. The inner batching is dynamically applied along $P_n$ as a complementary mechanism, which helps reduce the memory footprint when loading projections but does not reduce the volume memory footprint. Hence, through these two-level batchings, we achieve out-of-core FBP execution.


However, the outer batching produces larger projection data overlap, which introduces additional I/O overhead.
As shown in Fig.~\ref{fig:overlappeda}, the $i$-th outer batch $E^{i}$ requires more detector rows ($P_r$) for reconstruction and causes greater projection data overlap with the adjacent batch $E^{i+1}$.
Based on this observation, we design an efficient performance model-guided pipeline optimization to mitigate redundant I/O overhead.

\subsection{End-to-end Pipeline Optimization} \label{sec:Pipeline}

This section presents the end-to-end pipeline optimization in \method{}.
As shown in Fig.~\ref{fig:jetson_nano_pipeline}, we design a unified pipeline strategy for out-of-core FBP execution, leveraging \texttt{cudaStream} and \texttt{cudaEvent}. We adopt double buffering for the input projection data, enabling overlap of data loading and computation across batches. For the output volume, only a single buffer is maintained, as the data is written to storage immediately after computation. All CUDA kernels are launched asynchronously, with three key data dependencies: (1) The input projection must be loaded before computation, (2) The output volume must be computed before being written back to storage, and (3) Computation must occur after the previous volume data is written back to storage. 
To that end, the pipeline overlaps storing the output of the current batch $E^{i}$ with loading the next batch $E^{i+1}$ input. Within each outer batch, the computation of batch $F^{i}_{j}$ is overlapped with the data loading of the subsequent inner batch $F^{i}_{j+1}$.

\subsection{Model-guided I/O Optimization} \label{sec:Selection}
After converting FBP to out-of-core execution and applying our pipeline model, we can observe that the load, computation, and store operations involved in each batch vary and can overlap. 
Consequently, both the batch size and the scheduling order of batches are pivotal factors that determine the overall performance of \Xuetao{the} \method{} framework.

\subsubsection{Arithmetic Intensity Analysis} 
We conduct an Arithmetic Intensity (AI) analysis~\cite{Williams2009Roofline,HennessyPatterson2024} for $E^{i}$. The numerator represents the total number of floating-point operations (FLOPs) and the denominator reflects the amount of data movement in bytes, as shown below:
\begin{equation}\footnotesize
    AI = \frac{(\alpha{\cdot}P_{c}{\cdot}log(P_{c}){\cdot}{\Delta}{\cdot}{E^{i}}{\cdot}{P_n})_{\mathbf{FFT}} + (\beta\cdot\Omega{E^{i}}{\cdot}{V_y}{\cdot}{V_x}{\cdot}{P_n})_{\mathbf{BP}} } {(\Delta E^{i}\cdot P_{c}{\cdot}P_{n})_{\mathbf{Load}}   + (\Omega E^{i}{\cdot}V_y{\cdot}V_x)_{\mathbf{Store}} }.
    \label{eq:AI}
\end{equation}
Here, $\alpha$ and $\beta$ are constants representing the computational cost of the FFT (more details in Section~\ref{sec:cuFFT-filtering}) and back-projection operations, respectively. $\Delta E^{i}$ denotes the number of projection rows ($P_r$) involved, as determined by the overlaps in Fig.~\ref{fig:overlap}, $\Omega E^{i}$ denotes the corresponding volume of batch $E^{i}$. Since the FFT accounts for only a minor portion of the overall runtime, we consider only the FLOPs contributed by the BP stage.
Therefore, $AI$ may be written as
\begin{equation}\footnotesize
    AI = \beta\cdot P_n / {(\zeta+1)},
\end{equation}
where $\zeta = \frac{\Omega E^{i} \cdot V_y \cdot V_x}{\Delta E^{i} \cdot P_c \cdot P_n}$. It compares the volume size to the total projection size. The value of $\zeta$ determines whether the workload is \emph{I/O bound} or \emph{compute bound}, and consequently guides vendor optimization. Although the batch size of $E^{i}$ also affects the AI (i.e., larger batch sizes leading to higher AI), it is not the primary determining factor but serves as a tuning parameter to adjust the performance defined by $\zeta$.


\begin{figure}[t]
    \centering    
    \includegraphics[width=\columnwidth]{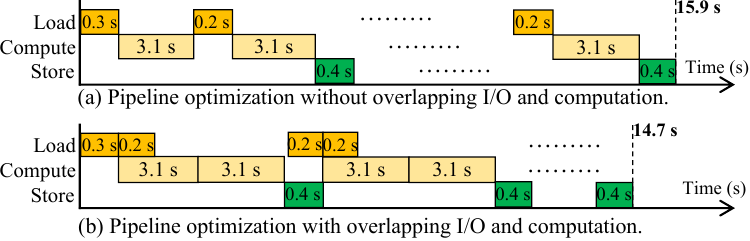} 
    \caption{End-to-end pipeline optimization. We adopt a double buffering approach using CUDA streams to hide the I/O overhead. This figure shows the pipeline timeline for Tomo\_30~\cite{de2018tomobank} dataset ($512^3$ output volume) on Jetson Nano. \textit{Load} and \textit{Store} indicate I/O, loading projections from storage and storing 3D volumes, respectively, while \textit{Compute} denotes filtering and BP. The variation in load time (0.2–0.3 s) is attributed to changes in the projection batch size. Pipeline overlap reduced runtime by 7.6\%, lowering runtime from 15.9~s to 14.7~s.
    }
    \label{fig:jetson_nano_pipeline}
\end{figure}
\subsubsection{Batch Size Optimization} \label{sec:BatchSize}
The selection of batch size is performed in two stages: (1) Determining whether the CT workload is compute-bounded or I/O-bounded, which guides the selection of the outer batch size of $E^{i}$. (2) Once the outer batch size is determined, the inner batch size of $F^{i}_{j}$ is selected based on the memory constraints of the edge device.
To illustrate this process, we discuss two scenarios in CT imaging: 
in the typical CBCT scenarios~\cite{scarfe2008cone,Jaffray2000FlatPanel}, the input and output sizes are comparable ($\zeta \approx 1$). According to Equation~\ref{eq:AI}, this leads to a relatively low AI, and the workload is consistently I/O-bounded. In such cases, maximizing AI becomes the primary objective. Therefore, selecting a larger $E^{i}$ is preferred to reduce data movement per FLOP. 
In contrast, micro‑CT~\cite{clark2014micro,compactCT,ying2017micro} scenarios generate massive output volumes from small projection inputs ($\zeta \gg 1$), resulting in a compute-bound workload where the bottleneck shifts to the back-projection (BP) stage. Under this condition, although a smaller $E^{i}$ causes additional I/O loading, it does not degrade overall performance due to the overlap enabled by our pipeline design. On the contrary, a smaller $E^{i}$ improves data locality, which in turn better supports the Runtime Kernels optimizations described in Section~\ref{sec:Implementation}. 
After $E^{i}$ is determined, the $F^{i}_{j}$  is in turn constrained to the hardware memory limit $Mem$, as expressed by the following inequality:
\begin{equation}\footnotesize
    \underset{\mathclap{\substack{\uparrow \\ \scriptsize \text{double} \\ \scriptsize \text{buffer}}}}{2} \cdot \Delta E^{i} \cdot \boxed{F^{i}_{j}} \cdot P_{c}  \cdot \underset{\mathclap{\substack{\uparrow \\ \scriptsize \text{FFT} \\ \scriptsize \text{complex}}}}{16} + \Omega E^{i} \cdot V_x \cdot V_y \cdot \underset{\mathclap{\substack{\uparrow \\ \scriptsize \text{float}}}}{4} < Mem
    \label{eq:mem}
\end{equation}
According to the geometry shown in Fig.~\ref{fig:overlap}, $\Delta E^{i}$ becomes smaller near the center and larger toward the periphery. This indicates that, under the condition of Equation~\ref{eq:mem}, the corresponding $F^{i}_{j}$ increases as $E^{i}$ approaches the center, resulting in reduced overlap near the projection center and allowing larger inner batch sizes.

\begin{figure}[t]
    \centering   
    \includegraphics[width=0.998\columnwidth]{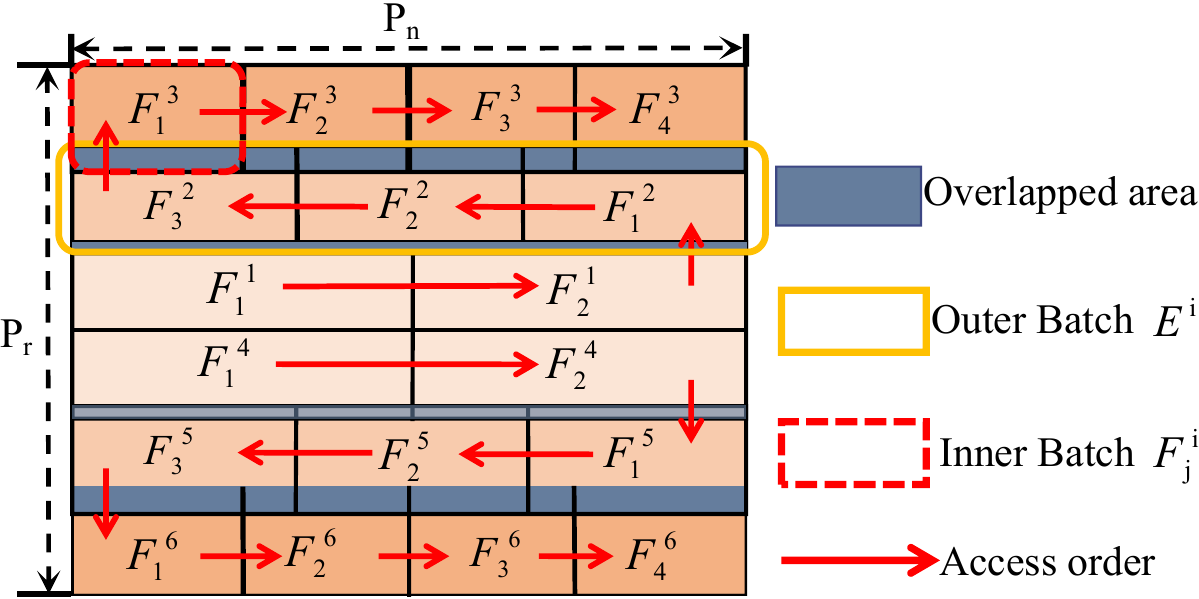} 
    \caption{Two S-shaped reorder path algorithm for the $P_n{\times}P_r$ input projections to enhance data locality. Two S-shaped traversals spread outward from the center, starting with little overlap that gradually increases.}
    \label{fig:loop_order}
\end{figure}
\subsubsection{Loop Reordering for Enhanced Data Locality}
As shown in Fig.~\ref{fig:loop_order}, two S-shaped traversal paths are adopted, starting from the center and progressing outward in both directions. 
With the S-shaped traversal, data reuse occurs between adjacent outer batches $E^{i}$ and $E^{i+1}$. In contrast, if each inner batch $F^{i}_{j}$  always starts from projection angle zero, such reuse cannot be achieved. 
The decision to traverse outward from the center is motivated by pipeline overlap considerations. Indeed, regions near the center are compute-bound with smaller $E^{i}$, while peripheral regions are I/O-bound with larger $E^{i}$. As described in Section~\ref{sec:Pipeline}, the implementation of the pipeline relies on double buffering. When processing from larger to smaller outer batches, the memory usage may exceed Equation~\ref{eq:mem}.
Meanwhile, the pipeline enables overlapping the next $E^{i+1}$ loading with the current $E^{i}$storing. 
By proceeding outward from the center, the pipeline effectively exploits this overlap, improving overall performance.


\section{GPU-optimized FBP Kernels} \label{sec:Implementation}

\begin{figure}[t]
   \centering
    \subfloat[$512^3$ output volume.]{
        \includegraphics[width=0.48\linewidth]{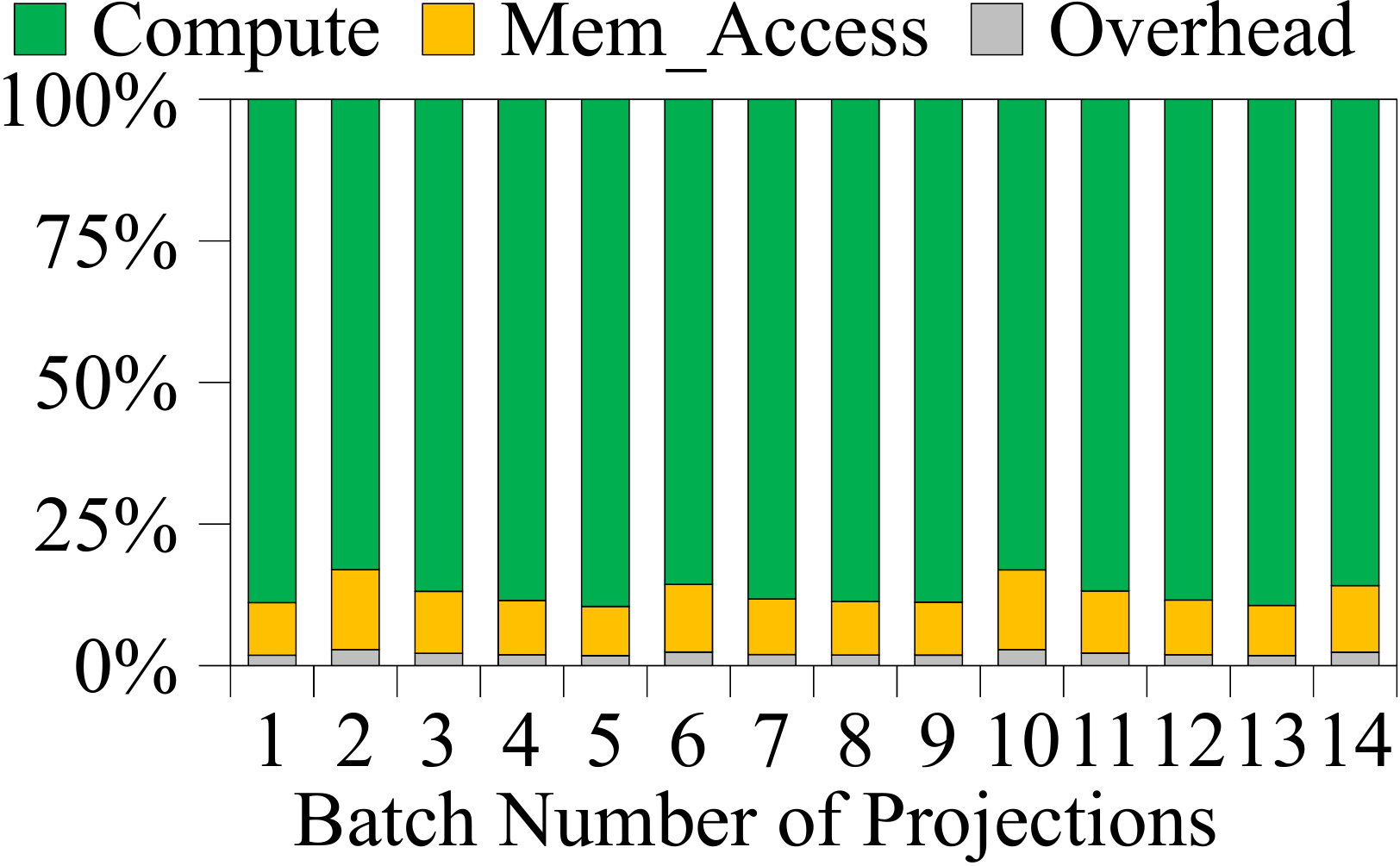}
        \label{fig:break1}
    }
    \subfloat[$1024^3$ output volume.]{
        \includegraphics[width=0.48\linewidth]{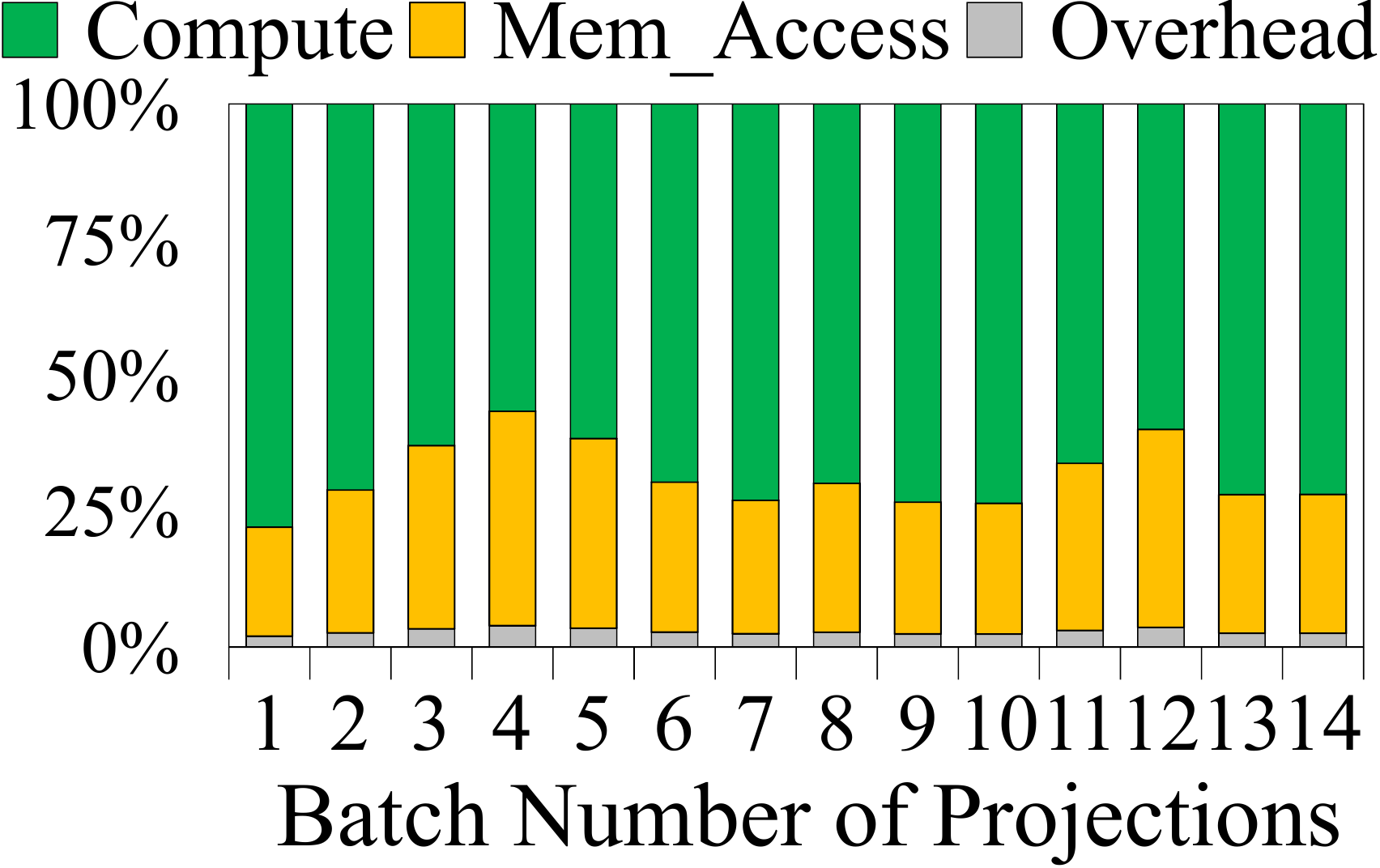}
        \label{fig:break2}
    }
    \caption{Computation breakdown of the BP kernel in RTK for Tomo\_29 reconstruction with resolution of (a)$512^3$ and (b)$1024^3$. \emph{Compute} means the compute occupancy. \emph{Memory\_Access} means memory occupancy. \emph{Overhead} means with thread assignment and other control operations.}
    \label{fig:breakdown}
\end{figure}

This section details the design of the Runtime Kernels in our \method{}, which addresses two critical bottlenecks in the back-projection process: (1) \emph{Memory access at those coordinates}, and (2) \emph{Computation of projection coordinates for each voxel.} 
In the baseline RTK~\cite{rit2014reconstruction} implementation, these two bottlenecks dominate the runtime, as shown in Fig.~\ref{fig:breakdown}. From the perspective of GPU global memory access, each voxel at every projection angle requires a coordinate computation followed by a GPU global memory access. This results in a fixed ratio between computation and memory access. However, the varying ratios shown in Fig.~\ref{fig:breakdown} arise from differences in the data locality of GPU global memory accesses across cases. We reduce and rearrange memory accesses, then we accelerate coordinate computation using TCs.

\subsection{\textit{cuFFT}-Accelerated Filtering Computation}\label{sec:cuFFT-filtering}
The filtering stage described in Equation~\ref{eqn:flt} and Fig.~\ref{fig:overview} applies a sharpening operation to each projection before back-projection. It forms an essential component of our GPU-optimized FBP Kernels. To reduce computational cost, we apply the Convolution Theorem~\cite{hirschman2012convolution} and perform the filtering in the frequency domain, where convolution becomes an element-wise multiplication. This is implemented using Nvidia’s \emph{cuFFT} library~\cite{NVIDIAcufft}, which provides highly optimized FFT primitives available for Jetson-based GPU architectures.
Unlike discrete GPUs such as the A100, Jetson devices employ a unified memory architecture in which the CPU and CUDA cores share the same physical memory. This removes PCIe transfer overhead, making memory bandwidth a critical resource. An efficient FFT-based filtering pipeline is essential to minimize memory traffic and sustain high throughput.

\subsection{Optimizing GPU Memory Access Pattern for BP} \label{sec:Reordering}

Efficient memory access is essential for achieving high throughput in the reconstruction process. In this section, we categorize and optimize the memory access patterns into the following types, with the overarching goal of minimizing expensive global memory transactions:
(1) \emph{Global Memory:} Data accessed for the first time, which must be fetched from global memory;
(2) \emph{L1 Cache:} Data accessed for the first time but already present in the L1 cache due to prior cache line placement;
(3) \emph{Shared Memory:} Data explicitly written to shared memory in advance and accessed from there.
We mitigate the cost of global memory accesses by improving data locality and reuse via on-chip caches and shared memory.

\subsubsection{GPU Memory Optimization Principles}

Both the L1 cache and shared memory are on-chip resources that are private to each GPU Streaming Multiprocessor (SM). This means data is not shared across SMs or kernel launches, and inter-kernel data reuse is not possible  (before the Hopper architecture). As a result, each kernel execution incurs at least one mandatory global memory fetch per access. 
Moreover, the CUDA warp stalls an instruction until all threads complete it. Consequently, if any thread accesses global memory, the entire warp suffers the memory latency, making intra-block data reuse critical.
This makes data reuse within a block especially important for achieving high performance.

\subsubsection{Optimal CUDA Thread/Block Configuration} \label{sec:block}

\begin{figure}[t]
    \centering
    \includegraphics[width=\columnwidth]{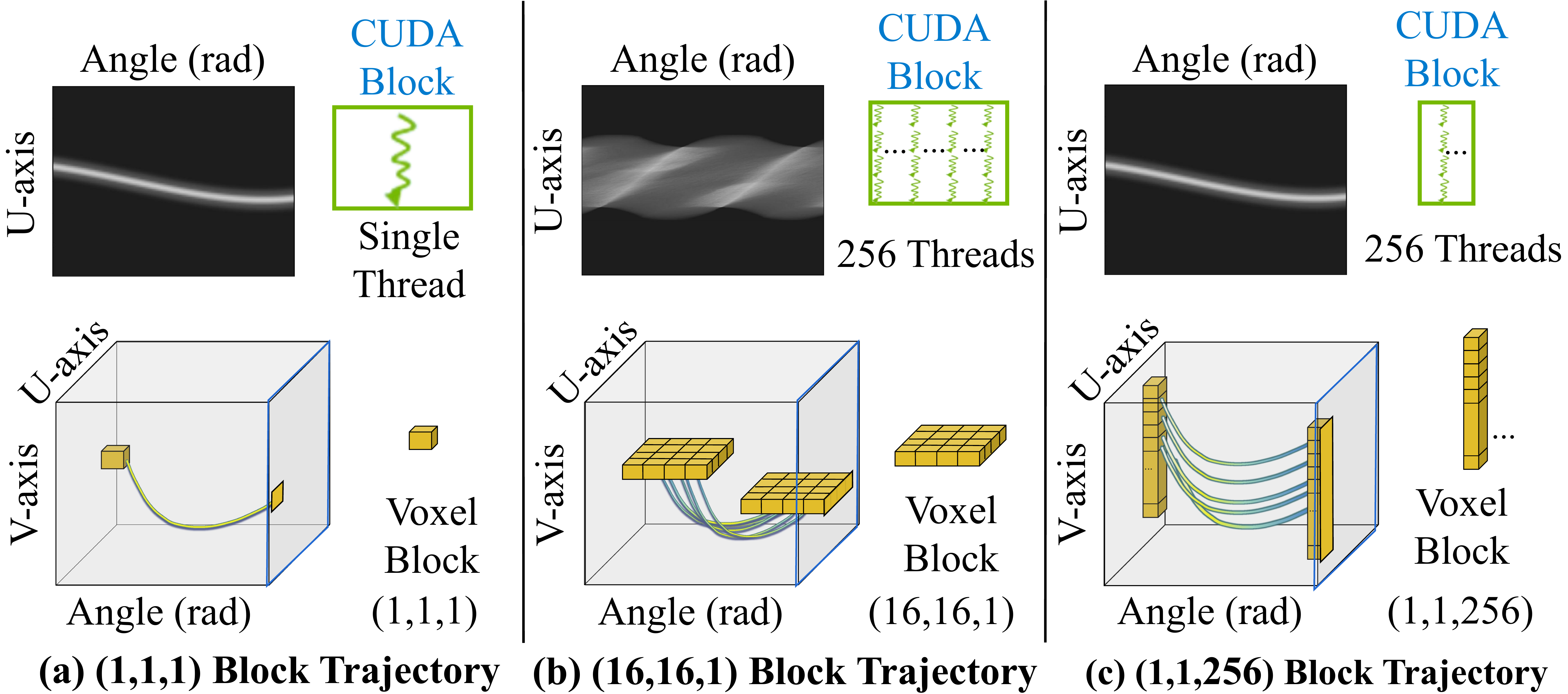} 
    \caption{Projection footprints for different voxel aggregation patterns. (a) Single voxel with a minimal footprint. (b) $16{\times}16{\times}1$ in-plane footprint expanding across the detector. (c) $1{\times}1{\times}256$ through-plane footprint expanding along the projection angle.
}
    \label{fig:mem_access}
\end{figure}

As shown in Fig.~\ref{fig:mem_access}a, computing a single voxel requires accessing its corresponding projection, which offers no data reuse and minimal data locality. Data reuse only emerges when computing multiple voxels within a thread block. We fix the voxel block size to 256 and compare two configurations: $(16,16,1)$ and $(1,1,256)$, as shown in Fig.~\ref{fig:mem_access}b and ~\ref{fig:mem_access}c. The data reuse occurs only along the XY-axes with irregular data locality, while the regular data locality is observed along the Z-axis. According to the Algorithm~\ref{alg:Block} (line 5-7), we map $(tx, ty, tz)$ voxels onto $(tx, ty, 1)$ thread block, using a loop of length $tz$ along the Z-axis within each thread to exploit data locality. 


\subsubsection{Lightweight Runtime Packing for Data Locality}
\begin{figure}[t]
    \centering
    \includegraphics[width=\columnwidth]{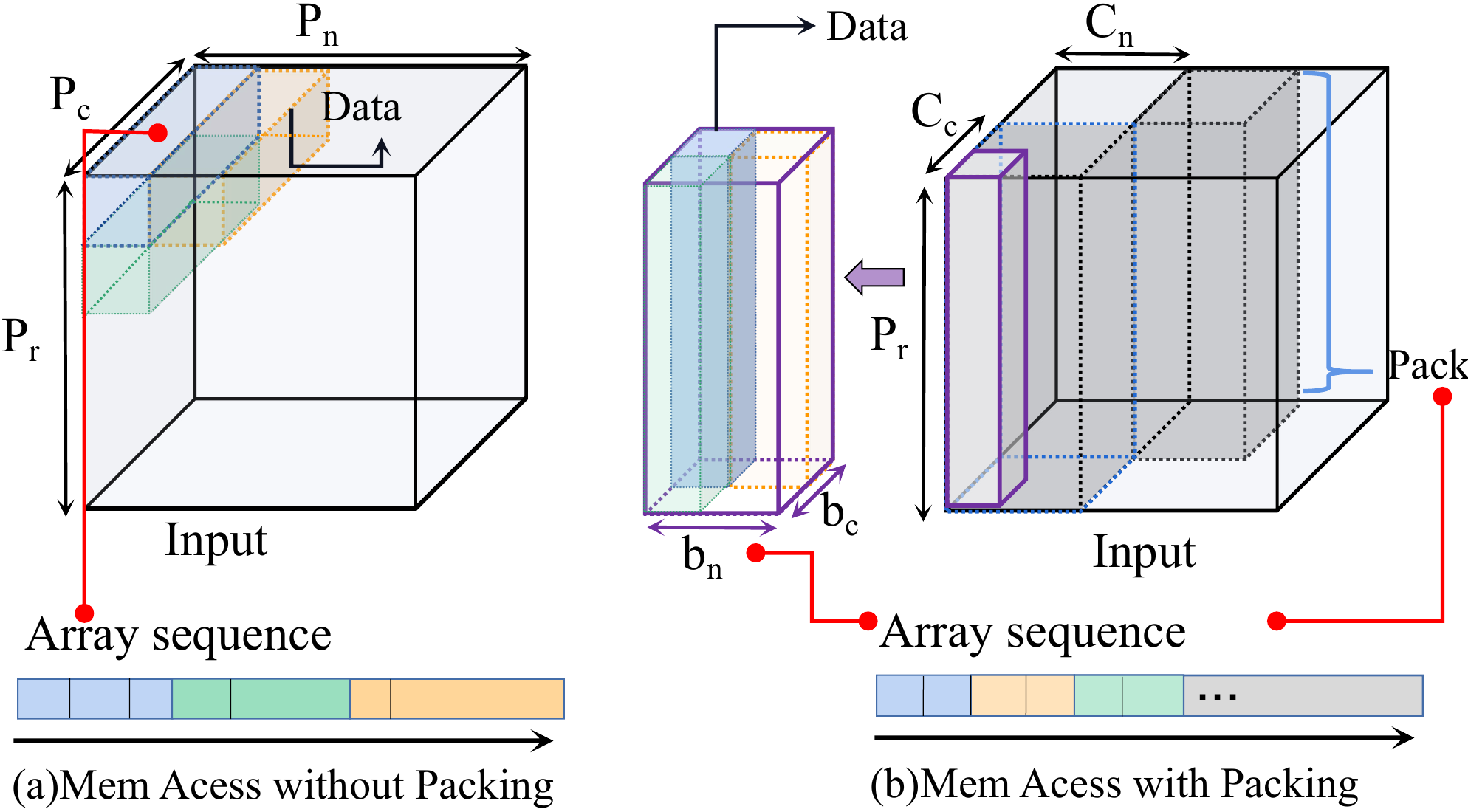} 
    \caption{Lightweight runtime packing to improve data locality.
(a) Original memory access pattern, where data are stored as \(P_n \to P_r \to P_c\). (b) Packed memory access pattern. The data are reorganized as \(C_n \to C_c \to b_c \to b_n \to P_r\): \(C_c\) and \(b_c\) form cache-friendly column groups, \(b_n\) groups eight projections for TCs computation, and the innermost \(P_r\) dimension enables contiguous access and more effective prefetching during interpolation.}
    \label{fig:packing}
\end{figure}

We perform projection data packing before back-projection to improve data locality.
To avoid introducing an additional preprocessing kernel, we fuse the packing operation with the write-back stage of the Filtering kernel, so that filtered projections are directly stored in the layout required by the back-projection stage. This design minimizes redundant global memory traffic and keeps the packing overhead to a minimum.
As shown in Fig.~\ref{fig:packing}, the original projection data are stored in the order \(P\_n{\rightarrow}P\_r{\rightarrow}P\_c\), where \(P\_n\), \(P\_r\), and \(P\_c\) denote the projection, row, and column dimensions, respectively.
We reorganize it into a packed format \(C_n{\rightarrow}C_c{\rightarrow}b_c{\rightarrow}b_n{\rightarrow}P_r\). The outermost dimension \(C_n\) is preserved from the original layout to keep low packing costs. 
The dimensions \(C_c\) and \(b_c\) divide the $P_c$ into smaller and cache-friendly groups to better match the memory access behavior of the \((tx,\,ty,\,1)\) thread block along the XY-plane. 
The \(b_n\) dimension is fixed to 8 (Section~\ref{sec:layout}), aligning with TCs requirements to enable eight coordinate computations to be processed concurrently. Compared with the original layout, this packed format enables more effective prefetching along \(P\_r\), improving memory locality during back-projection. The length of \(P\_r\) is associated with the outer batch \(E^{i}\), which is determined by the CBCT configuration in Section~\ref{sec:BatchSize}.

\subsubsection{Shared Memory Cache Optimization}
In image reconstruction, the range of computed coordinate indices is known and bounded, which allows us to design a more efficient caching strategy. We implement a shared memory cache, removing the need for a hash table or other associative lookup structures. When a data element is fetched from global memory, we update the cache tag and fetch the data to the corresponding shared memory cache line. Then the thread can check the cache tag and retrieve the data directly from shared memory, effectively replacing an expensive global memory access with two low-latency shared memory operations.




\input{Algorithms/Block}

\subsection{Optimizing BP Computation using TCs} \label{sec:Tensor Core}
We optimize the BP bottleneck (Fig.~\ref{fig:breakdown}) via MMA-mapped $M_{\psi}\cdot[i,j,k,1]^T$, with accuracy by decomposition and scaling. 
Nvidia provides a TC programming interface through MMA (matrix-multiply-accumulate) PTX instructions, e.g., \emh{mma.sync}, which execute fused \emph{A${\times}$B+C} operations on small per-warp matrix tiles, enabling efficient mixed-precision computation~\cite{Nvidia_ptxisa_contents,Nvidia_ptxisa_mma} for theoretically 8$\times$ speedup over CUDA cores (Tab~\ref{tbl:jetson}). This advanced architecture can bring the opportunity to accelerate the major bottleneck (i.e., projection coordinate computation) in the GPU-optimized FBP Runtime Kernels. However, achieving high performance requires careful control of data layout and numerical stability.


\subsubsection{MMA Primitive for BP Computation}
We first minimize the computational workload in our implementation before leveraging TCs. In back-projection, the coordinate computation involves mapping each voxel's 3D coordinates onto the 2D projection plane. As shown in the geometric analysis in Section~\ref{sec:block}, when the XY-axes of the voxel are fixed, its horizontal coordinate on the projection plane remains constant. Therefore, as presented in Algorithm~\ref{alg:Block} line (5-7), each thread in a (tx, ty, 1) block can reuse the same XY-plane coordinate computation across all tz iterations along the Z-axis, thereby decreasing arithmetic cost.
To utilize TCs, our implementation directly invokes MMA PTX instructions~\cite{Nvidia_ptxisa_contents,Nvidia_ptxisa_mma}. This low-level approach enables fine-grained control, preserving the logic structure of the original CUDA implementation and facilitating a seamless transition to TCs acceleration.

\subsubsection{Optimizing BP Computation with FP16 TCs} \label{sec:layout}

\begin{figure}[t]
    \centering
    \includegraphics[width=0.95\columnwidth]{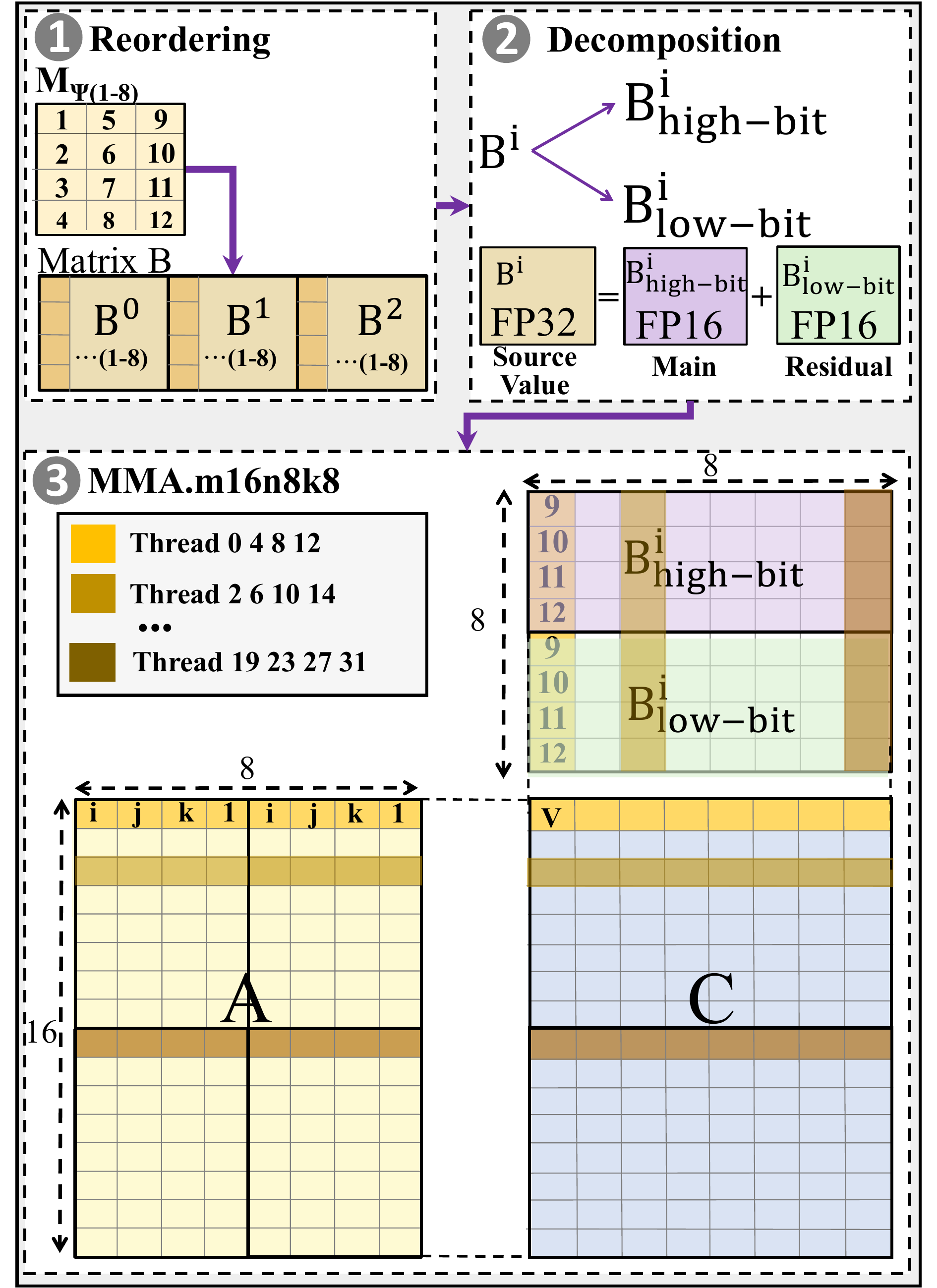} 
    \caption{Optimizing coordinate calculation in BP kernel (Alg.~\ref{alg:BP}, lines 6–7) using FP16 TCs with the \emph{m16n8k8} MMA instruction.
    (1) Columns 1–3 of each $M_{\psi(1\text{-}8)}$ group are first reorganized into three $4\times8$ blocks $B^{0,1,2}$. 
    (2) We devised an efficient FP32-to-FP16 decomposition strategy to split each FP32 $B^i$ into two $4\times8$ FP16 high-bit and low-bit sub-matrices i.e., $B_{high-bit}^i$ and $B_{low-bit}^i$.
    (3) The FP16 mma.m16n8k8 instruction is launched to accelerate the multiplication with projection matrix $A$ and each $B^{i}_{\text{high/low-bit}}$.
    }
    \label{fig:fp16}
\end{figure}

\input{listing/decomposition}

    




Although the TCs promise significant speedups, three obstacles arise when applying them to coordinate computation:
{
(1) \emph{Data layout.} FP16 TCs impose strict requirements on fragment layout, such as \emph{m8n8k4} and \emph{m16n8k8}; 
(1) \emph{Precision.} The TCs operate in FP16 format (1-bit sign, 5-bit exponent, 10-bit mantissa), which requires redesigning the matrices to decompose FP32 numbers into components suitable for processing. 
(3) \emph{Dynamic range.} The dynamic range of FP16 is narrower than that of FP32, making potential overflow errors during computation.
}

In our case, the coordinate computation $K$-dimension is originally fixed at $k=4$, which limits the effective use of MMA instructions.
The \emph{m8n8k4} and \emph{m16n8k8} layouts represent different TCs fragment layouts, each with varying levels of hardware support and performance characteristics.
The \emph{m8n8k4} layout was the earliest format supported by first generation TCs and remains available on Ampere GPUs. However, it cannot utilize the theoretical peak performance on Ampere, due to its suboptimal scheduling on modern warp-level execution units. In contrast, \emph{m16n8k8} is the default layout used by \texttt{mma.sync} instructions workloads internally optimized for Ampere architecture, aligning with warp-level fragment sizes for higher instruction-level parallelism and minimal pipeline stalls.


To leverage the \emph{m16n8k8} layouts while maintaining accuracy, we decompose the projection matrices into high-order and low-order components, as described in Fig.~\ref{fig:fp16}. The high-order component corresponds to the FP16 truncation of the original FP32 value, while the low-order component is then computed as the residual difference between the original value and this high-order component, as in Listing~\ref{listing:FP32toFP16}.
In addition to precision concerns, we also address the overflow errors of the FP16 precision. 
In our design, the projection matrix is scaled by ${1}/{L_{SD}}$ before matrix operations, and the resulting coordinates are rescaled by ${L_{SD}}/{z}$ after the matrix operations. This normalization keeps intermediate values within the FP16 range avoiding overflow.



\section{Evaluation} \label{sec:evaluation}
\begin{figure}[t]
    \centering
    \begin{minipage}{0.475\columnwidth}
        \centering
        \subfloat[Shepp-logan~\cite{feldkamp1984practical} result.]{
            \includegraphics[width=0.925\linewidth]{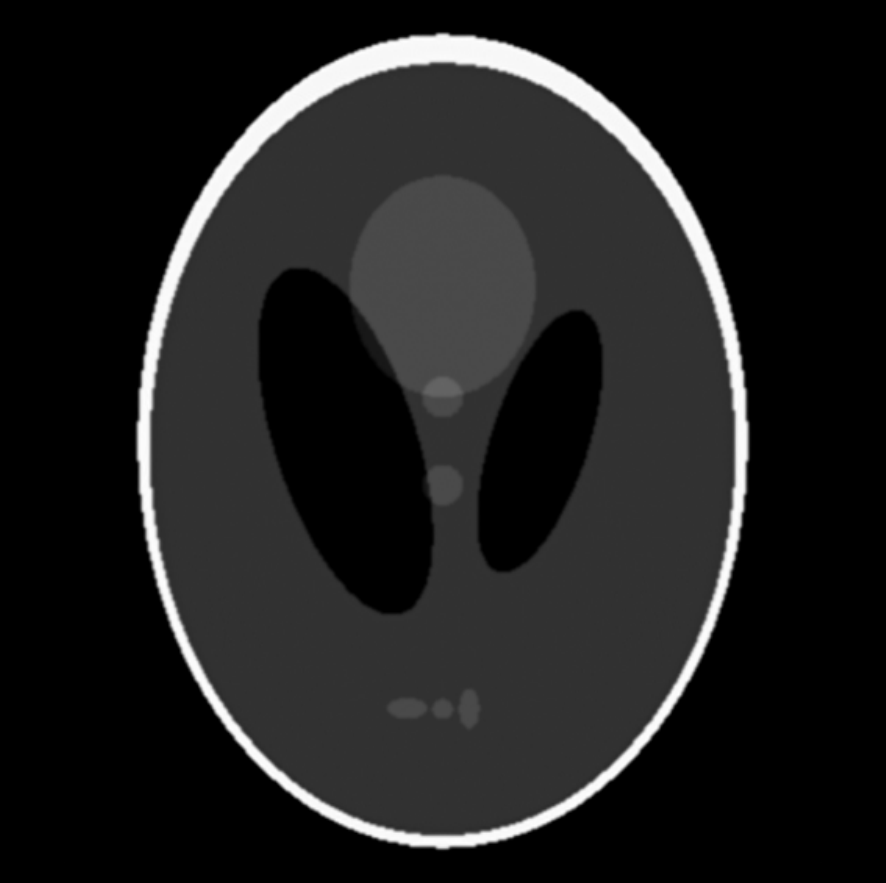}
            \label{fig:reconstructiona}
        }
    \end{minipage}
    \begin{minipage}{0.475\columnwidth}
        \centering
        \subfloat[Tomo\_30~\cite{de2018tomobank} result.]{
            \includegraphics[width=0.925\linewidth]{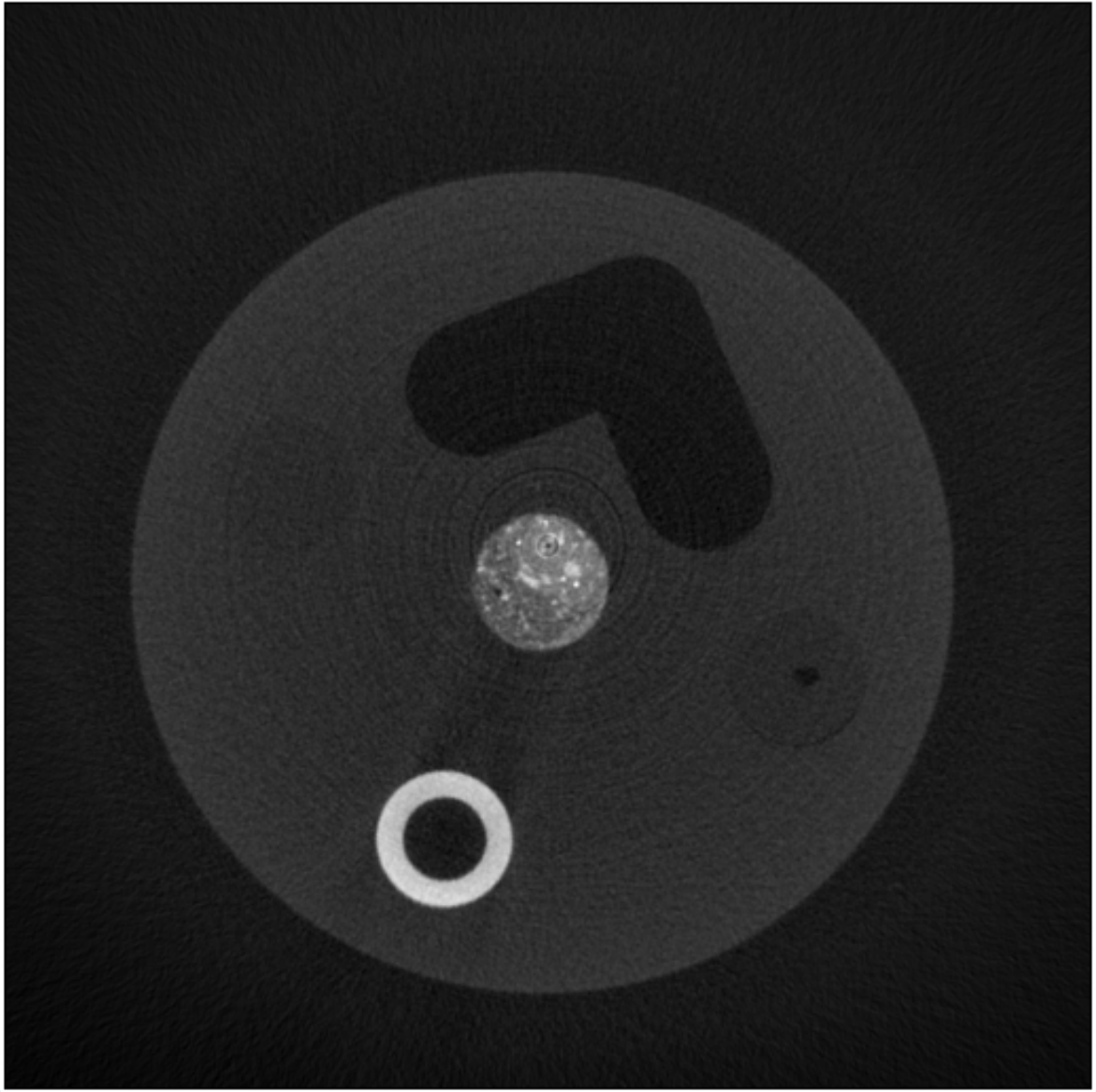}
            \label{fig:reconstructionb}
        }
    \end{minipage}
    
    \caption{Reconstruction results of \method{} for the Shepp-Logan and Tomo\_30 datasets at a resolution of $2048^3$. 
    }

    \label{fig:reconstruction}
\end{figure}

We implemented \method{} in Nvidia CUDA and evaluated it on datasets such as the Shepp-Logan phantom~\cite{Shepp1974Fourier} and the TomoBank dataset~\cite{de2018tomobank} on Nvidia edge platforms.
Fig.~\ref{fig:reconstructiona} and Fig.~\ref{fig:reconstructionb} show reconstruction slices for the Shepp-Logan phantom and the Tomo\_30 dataset at $2048^3$, confirming that \method{} preserves reconstruction quality.


\input{tables/Dataset}
\subsection{Datasets and Evaluation Environment}

{\bf{Datasets.}} As shown in Table~\ref{tbl:datasets}, we use two real-world datasets using industrial scanners to conduct the evaluations: (1) \emph{TomoBank dataset} with
four sub-datasets: bone local (Tomo\_27), bone local stone (Tomo\_28),
candie local (Tomo\_29), smiling sample (Tomo\_30)~\cite{de2018tomobank}, and (2) \emph{Bumblebee dataset}, which is scanned by a Nikon Metrology HMX ST 225 micro-CT scanner.
The projection parameters were: $L_{sd} = 672.5$, $L_{so} = 39.8$, $P_c = P_r = 2000$, $\lambda_c = \lambda_r = 0.2$ and $P_n = 3142$ (projections acquired $P_n = 6401$ total). For each TomoBank dataset, we evaluate performance by scaling the output volume size from \(256^3\) up to \(3072^3\).


{\bf{Evaluation environment.}} Our evaluation spans three representative deployment scenarios: a high-end Nvidia DGX system equipped with 8$\times$A100 80GB GPUs, an edge AI module featuring Jetson AGX Orin (64GB); and a Jetson Orin Nano Super (8GB). All platforms are operated on Linux with CUDA 12.2+, sharing a unified software stack compiled with architecture-specific flags: \texttt{-arch=sm\_80} for Ampere (A100) and \texttt{-arch=sm\_87} for Orin architectures.

\input{tables/Performance}

{\bf{Accuracy validation.}} We conduct a voxel-wise comparison RTK~\cite{rit2014reconstruction} to evaluate the reconstruction accuracy of \method{}.
The RMSE remains below $1{\times}10^{-4}$ for all voxels, which is sufficient to meet XCT imaging quality requirements, since XCT projections are typically acquired with 16-bit precision and reconstructed volumes generally require no more than a 16-bit dynamic range when expressed in Hounsfield units~\cite{schreiber2011hounsfield}.
These results demonstrate that \method{} can guarantee the reconstruction quality.


\begin{figure}
    \centering
    \includegraphics[width=\columnwidth]{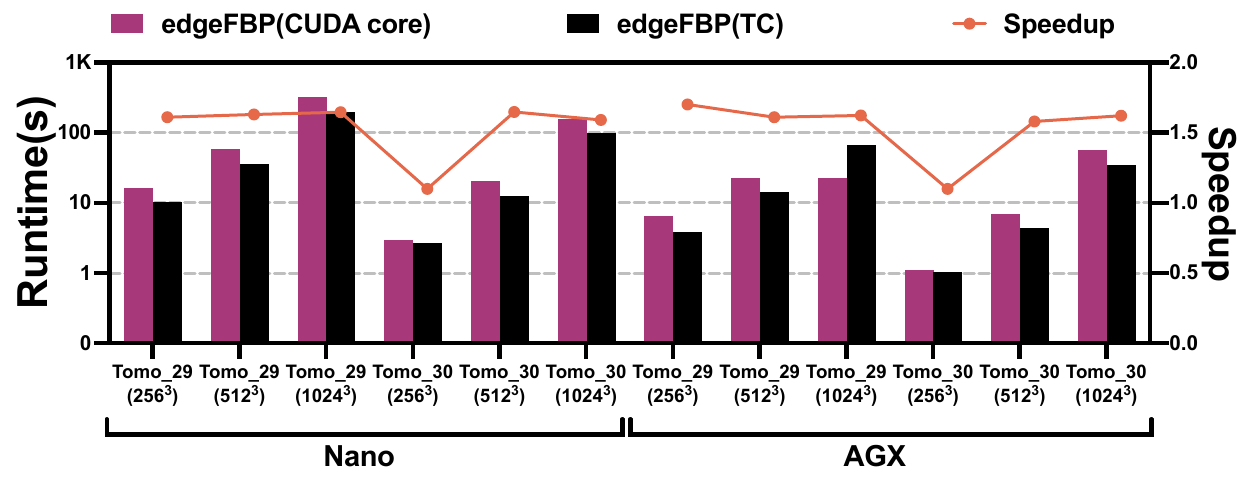}
    \caption{Performance evaluation of the TC-accelerated BP kernel on Orin Nano and AGX Orin. The optimized kernel achieves up to 1.53$\times$ and 1.54$\times$ speedup, respectively.}

    \label{fig:TCresult}
\end{figure}

\begin{figure}[t]
    \centering
    \includegraphics[width=\columnwidth]{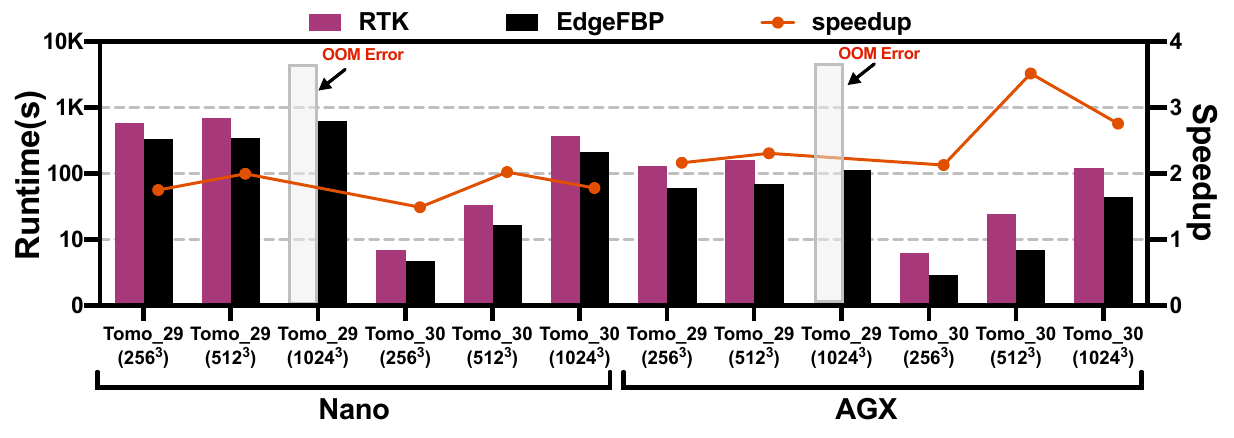} 
    \caption{Detailed end-to-end performance improvement of \method{} on Jetson Orin Nano under growing output.
    "OOM" = "Out of Memory". Compared to the baseline, \method{} achieves up to 1.83$\times$ and 2.57$\times$ speedup on Jetson Orin Nano and Jetson AGX Orin, respectively.
    }
    \label{fig:nano_scalability}
\end{figure}

\begin{figure}
    \centering
    \includegraphics[width=\columnwidth]{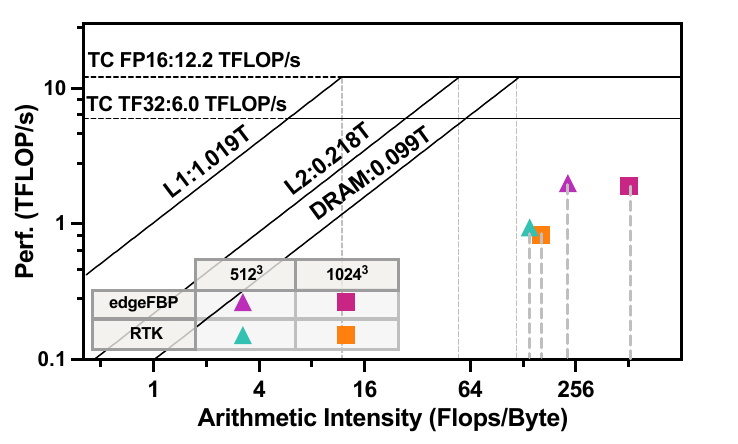} 
    \caption{Roofline analysis of the BP kernel on Jetson Nano using the Tomo\_30 dataset at $512^{3}$ and $1024^{3}$ resolutions. Results are collected using the Nvidia Nsight Compute (NCU) profiler~\cite{iyer2016gpu}. The optimized BP kernel achieves 1.49$\times$--2.15$\times$ speedup over the baseline.}
    \label{fig:roofline}
\end{figure}

\subsection{Throughput Performance Evaluations}
We conduct end-to-end evaluations to measure the performance of the \method{} on Jetson Nano and AGX. Detailed results are presented in Table~\ref{tbl:performance} and Fig.~\ref{fig:nano_scalability}, where \emph{Full Size} and the \emph{Max Batch} in the table denote the maximum memory capacity and the batch capacity of each dataset.
Despite the strict power and memory constraints of edge platforms, \method{} consistently delivers strong performance, achieving speedups of up to $1.83\times$ on the Jetson Nano and $2.57\times$ on the Jetson AGX compared to the RTK baseline. 
Moreover, the proposed optimization allows \method{} to handle large-scale reconstruction workloads on the Jetson platform with the SoC chip feature and the limited memory capacity, while the RTK implementation fails due to out-of-memory (OOM) errors for large volumes.
These gains demonstrate the effectiveness of our framework-level optimizations as well as the GPU-optimized FBP kernel design, both of which are critical for sustaining high-throughput CT reconstruction on resource-constrained edge devices.

We further evaluate the performance of the BP kernel in \method{} in Fig.~\ref{fig:TCresult}. The results show that the TCs optimized BP kernel achieves speedups of approximately $1.53\times$ and $1.54\times$ on the Jetson Orin Nano and Jetson AGX Orin platforms, respectively. These results demonstrate that the proposed TC-optimized kernel effectively improves computational efficiency and significantly enhances the throughput of the BP computation.

\subsection{Roofline Analysis of the BP Kernel}
To evaluate the efficiency of BP kernel, we perform roofline analysis on the Jetson Nano GPU using FP16 TCs (peak: $12.2$ TFLOPS). As shown in Fig.~\ref{fig:roofline}, both \method{} and RTK reach arithmetic intensities above the bandwidth ceilings, indicating they are largely compute-bound. Moreover, \method{} attains higher effective throughput than RTK, yielding speedups ranging from 1.49$\times$-2.15$\times$ across datasets (average 1.83$\times$). These results indicate efficient use of TCs and scalable performance with increasing dataset size.

\subsection{Out-of-core Reconstruction Capability}
Table~\ref{tbl:performance} demonstrates the out-of-core capability of \method{} on Jetson Nano and Jetson AGX when reconstructing $2048^3$ and $3072^3$ volumes. 
For example, the memory footprints of $2048^3$ and $3072^3$ volumes are 32\,GB and 108\,GB, respectively, which significantly exceed the memory capacity of Nano and AGX according to the specifications in Table~\ref{tbl:jetson}.
In these cases, the RTK baseline fails with out-of-memory errors, while \method{} successfully completes the reconstruction. In summary, these results denote that the projection and volume partition strategy in Section~\ref{sec:Out-of-Core} enables practical out-of-core reconstruction on memory-constrained edge platforms.


\begin{figure}[t]
    \centering
    \includegraphics[width=\columnwidth]{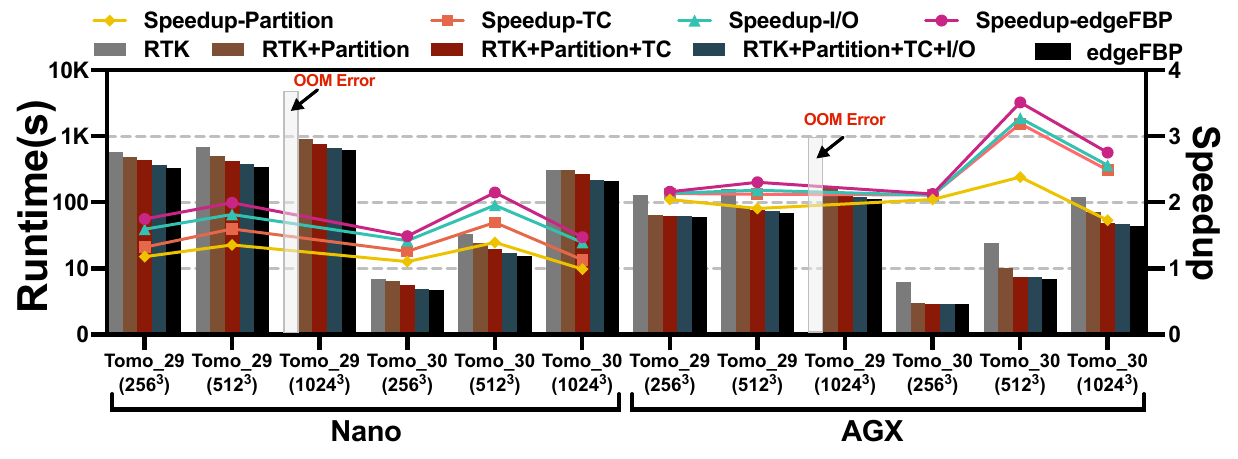} 
    \caption{Step-wise breakdown evaluation \Xuetao{of the optimized FBP implementations} on Jetson Nano and AGX.
        \textit{RTK}~\cite{rit2014reconstruction} is baseline CUDA implementation, \textit{TC} denotes the Tensor Core acceleration, and \textit{Batch} applies model-guided storage I/O optimization.
        \method{} integrates all three optimizations.
        All speedups are relative to \textit{RTK}.
        }

    \label{fig:Compare}
\end{figure}

\subsection{Evaluation of the Proposed Optimizations} \label{sec:breakdown}

As illustrated in the framework design, we mainly propose projection \& volume partition (partition), TC-accelerated BP computation (TC), model-guided I/O optimization (I/O), pipeline optimization (edgeFBP), and memory-access pattern optimization.
We examine the effect of step-wise optimization through a sequence of experiments in Fig.~\ref{fig:Compare}.

\begin{figure}[t]
    \centering    
    \includegraphics[width=\columnwidth]{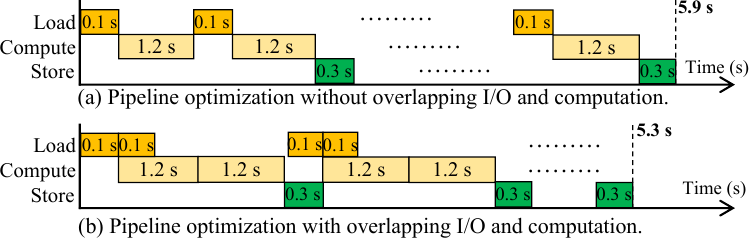} 
    \caption{
    Pipeline timeline evaluations on the Tomo\_30 dataset ($512^3$ output) on Jetson AGX.
    Pipeline overlap reduced runtime by 10.36\%, lowering runtime from 5.9~s to 5.3~s.
    }
    \label{fig:jetson_agx_pipeline}
\end{figure}
\subsubsection{Evaluation of the Projection \& Volume Partition}
After the partition, our framework effectively alleviates the OOM issue for Tomo\_29 ($1024^3$) dataset on Jetson Nano and AGX platforms. Meanwhile, it achieves 1.20$\times$ and 2.02$\times$ speedup for the rest datasets on both platforms. These results highlight the effectiveness of the partitioning strategy in enabling out-of-core imaging by addressing OOM issues and facilitating efficient overlap between I/O and computation.

\subsubsection{Evaluation of TCs Optimization}
Compared to the baseline RTK, using TCs can achieve around 1.40$\times$ and 2.41$\times$ speedup on Jetson Nano and AGX platforms (without Tomo\_29 ($1024^3$)). Compared to the Partition approach, using TCs can achieve 1.17$\times$ and 1.22$\times$ on both platforms. Meanwhile, the TC-accelerated BP kernel achieves about 1.53$\times$ and 1.54$\times$ speedup compared to the CUDA-based BP kernel in \method{}. These results denote that the proposed TCs acceleration can significantly enhance the BP throughput.



\subsubsection{Evaluation of Model-guided I/O Optimization}
After the batched I/O optimization, the framework achieves 1.64 $\times$ and 2.45 $\times$ speedup on Jetson Nano and AGX compared to the baseline RTK approach. Compared to the Partition approach, the framework achieves 1.37$\times$ and 1.25$\times$ acceleration on both platforms. The above results denote that the proposed batch size optimization can effectively reduce the data movement and improve the data locality.


\subsubsection{Evaluation of Pipeline Optimization}
With end-to-end pipeline optimization, \method{} achieves speedups of 1.77$\times$ on Jetson Nano and 2.57$\times$ on Jetson AGX over the baseline. Compared to the partition-based approach, it delivers additional speedups of 1.47$\times$ and 1.32$\times$, respectively. Pipeline timeline analyses on Jetson Nano and AGX (Figs.~\ref{fig:jetson_nano_pipeline} and \ref{fig:jetson_agx_pipeline}) show that the proposed optimizations effectively eliminate pipeline bubbles by overlapping I/O and computation, leading to higher pipeline utilization.



\begin{figure}[t]
    \centering
    \includegraphics[width=\columnwidth]{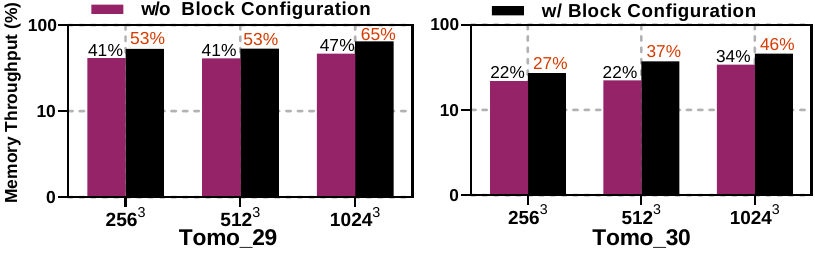} 
    \caption{performance evaluation of memory-access optimization in BP kernel on Jetson Nano through optimal CUDA block configuration. The packed GPU memory-access pattern improves performance by 11\% and 14\% on the Tomo\_29 and Tomo\_30 datasets, respectively.
    }
    \label{fig:mem_access_result}
\end{figure}

\subsubsection{Evaluation of Memory-Access Pattern Optimization}
We further evaluate the performance of memory-access pattern optimization. Specifically, we select the Jetson Nano platform with lower memory bandwidth to compare the memory throughput using Nsight Compute. As shown in Fig.~\ref{fig:mem_access_result}, the \method{} increases 11\% and 14\% of the memory throughput for Tomo\_29 and Tomo\_30 datasets. These results denote that the proposed configuration with Z-axis locality $(1,1,256)$ consistently outperforms the baseline $(16,16,1)$ across all datasets and output sizes, guaranteeing the effectiveness of our data locality design.


\subsection{Power Efficiency Comparison: Jetson vs. DGX}

\begin{figure}[t]
    \centering
    
    \centering
    \includegraphics[width=1.04\columnwidth]{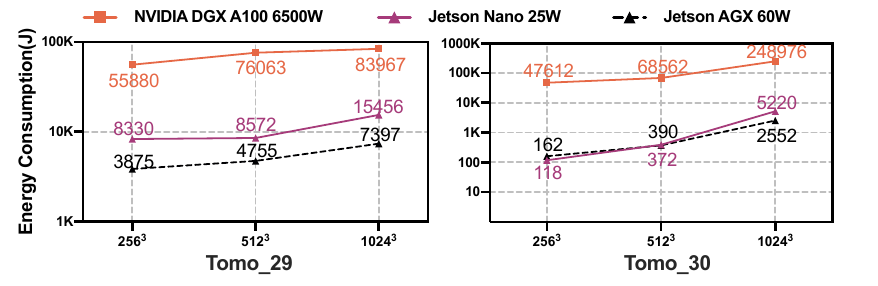} 
    \caption{Comparison of energy consumption across Jetson Nano, Jetson AGX, and a DGX A100 system (eight Nvidia A100 GPUs), along with the end-to-end evaluation of \method{} on the Tomo\_29 and Tomo\_30 datasets. Jetson AGX Orin and Jetson Nano achieve up to $18\times$ and $48\times$ higher energy efficiency than the DGX A100 system, respectively.}
    \label{fig:energynano}
\end{figure}

Fig.~\ref{fig:energynano} compares the energy efficiency of Jetson Nano and Jetson AGX with an Nvidia DGX A100 system equipped with $8\times$A100 80GB GPUs.
To evaluate energy efficiency, we measure and compare power consumption across these platforms. Jetson AGX is significantly more energy efficient than the DGX A100 system, achieving $18\times$ to $27\times$ higher efficiency for output sizes from $256^3$ to $1024^3$.
Jetson Nano further improves energy efficiency, achieving $5{\times}{\sim}9{\times}$ on Tomo\_29 and up to $48\times$ on Tomo\_30 with $1024^3$ output.
These results highlight the Jetson series as low-power alternatives to conventional HPC systems for imaging.
Note that we do not directly compare Jetson with discrete GPUs, such as the A100 or lower-end models. Jetson is an integrated SoC platform that integrates the CPU, GPU, memory, and storage in a single device, whereas GPU operates only as an accelerator and requires a separate host system that also consumes power.



\subsection{Impact of \method{} on Real-World Applications}
Jetson devices are compact SoC platforms and are cheaper than datacenter GPU systems. For example, Jetson developer kits typically cost a few hundred to a few thousand USD~\cite{NVIDIA2025JetsonPurchase}, whereas an Nvidia DGX system with eight A100 GPUs usually exceeds 100,000 USD~\cite{NVIDIA2024DGXA100}. As a result, a CT reconstruction system built on Jetson SoCs using \method{} can be one to two orders of magnitude cheaper than DGX-based solutions.
Beyond cost, \method{} enables high-quality CT reconstruction on low-power and compact platforms. Portable and point-of-care scanners often cannot accommodate high-end GPUs, but \method{} allows reconstruction to run locally on Jetson modules within a 10$\sim$40~W power budget. \method{} enables large-scale reconstructions when device memory is limited.
\method{} is also well suited for robotic and mobile imaging systems~\cite{salcudean2022robot}, enabling high-quality, local CT reconstruction on compact, low-power platforms.
Since Jetson modules are widely used in such platforms, \method{} integrates naturally and enables fast on-device 3D reconstruction. 
By leveraging Tensor Cores through a hardware-aware formulation, \method{} achieves high energy efficiency under strict power constraints, making it suitable for portable, embedded, and always-on imaging applications.

\section{Conclusion}\label{sec:conclusion}
We present \method{}, an efficient out-of-core CT reconstruction framework on Jetson platforms. By combining memory-efficient out-of-core partitioning with Tensor Core-accelerated BP and memory-access optimization, \method{} enables large-scale 3D reconstruction under the limited memory and power budget of embedded GPUs. Our evaluation shows that \method{} outperforms the widely used RTK library by up to $1.83\times$ on Jetson platforms and achieves up to $48\times$ higher energy efficiency compared with DGX-class systems. These results demonstrate the feasibility of high-quality tomographic imaging on edge platforms, enabling portable medical, industrial, and scientific imaging applications.




%% file: tables/JetsonDevicesTable.tex
\setlength{\tabcolsep}{1pt} 

\begin{table}[t]
    \centering
    \caption{
     Motivation of using Jetson series for CT imaging, emphasizing their power efficiency, integrated SoC design, and balanced performance. \emph{TC} refers to Tensor Cores.
    }

        
    \resizebox{\linewidth}{!}
    {
        \begin{tabular}{|c|c|c|c|c|c|c|c|c|c||c|c|}
        \hline
            \multicolumn{2}{|c|}{ \textbf{Jetson Series}} &  \multicolumn{3}{c|}{\textbf{AGX Orin}} & \multicolumn{2}{c|}{\textbf{Orin NX}} & \multicolumn{3}{|c||}{\textbf{Orin Nano}}   & \textbf{A100} & \textbf{H100}\\ \hline
             \multicolumn{2}{|c|}{ Module Type}  & 64GB & Industrial & 32GB & 16GB & 8GB & \textbf{Super} & 8GB & 4GB & - & - \\ \hline
        
        \multirow{2}{*}{GPU}&  Tensor Cores & \multicolumn{2}{c|} { 64} &  56 & \multicolumn{2}{c|}{  32 } &\multicolumn{2}{c|}{ \textbf{ 32} }  &   16  & \textbf{ 512} &  528   \\ \cline{2-12}
                & CUDA Cores& \multicolumn{2}{c|} { 2048 } & 1792 & \multicolumn{2}{c|}{ 1024} & \multicolumn{2}{c|}{\textbf{1024}}   & 512 & \textbf{8192} & 16896 \\ \hline
        
        \multirow{2}{*}{ Perf. (TFLOPS)} & \textbf{ FP16 w/ TC} &  43  &  39  &  27   &  19  &  19  & \textbf{ 17}  &   17  &   8.5  & \textbf{ 312} &  989.4  \\ \cline{2-12}
                & FP32 w/o TC & 5.3  & 4.8  & 3.8   & 2.4  & 2.4  & \textbf{2.1}  & 2.1  & 1.04  & \textbf{19.5} & 66.9 \\ \hline
        
        \multicolumn{2}{|c|}{   CPU Arm v8.2 Cores} & \multicolumn{3}{c|} { 12} &   8  &   6  & \multicolumn{3}{c||} {   6 } & {   -} & {   -} \\ \hline
        \multicolumn{2}{|c|}{Memory (GB)} & \multicolumn{2}{c|} {64 } & 64  & 32  & 16  & \multicolumn{2}{c|} {\textbf{8} } & 4 &\textbf{80} & 80 \\ \hline
        
        \multicolumn{2}{|c|}{ \textbf{ Power (Watt)}} & 
         15$\sim$60 & 
          15$\sim$75 & 
          15$\sim$40 &
         \multicolumn{2}{c|} { 
          10$\sim$15$\sim$25$\sim$40} & 
         \multicolumn{3}{c||} {
         \textbf{ 7$\sim$15$\sim$25}} & \textbf{ 400} &  700  \\ \hline
        \end{tabular}    
    }

    \label{tbl:jetson}
\end{table}

%% file: tables/Parameters.tex
\begin{table}[t]
\centering
\caption{Definitions of the geometric parameters in the CBCT imaging system.}

\resizebox{\linewidth}{!}
{
    \begin{tabular}{|c|c|c|}
    \hline
    \textbf{ Symbol} & \textbf{ Description} & \textbf{ Unit} \\ \hline
    $P_n$  & The number of views & — \\ \hline
     $P_c, P_r$ & Columns, rows of a projection in U-, V-axis & pixel \\ \hline
    $V_x, V_y, V_z$ & The number of voxels in X-, Y-, Z-axis & voxel \\ \hline
    $\psi$  & Rotation radian (step is $2\pi/P_n$) & rad \\ \hline
    $M_{\psi}$ & A projection matrix of size 3$\times$4 at radian $\psi$ & — \\ \hline
    $\lambda_c, \lambda_r$ & Pixel pitch at U- and V-axis & mm/pixel \\ \hline
    $\lambda_x, \lambda_y, \lambda_z$ & Voxel pitch at X-, Y-, and Z-axis & mm/voxel \\ \hline
    $\omega_c, \omega_r, \omega_{cor}$ & Offset of Detector at U- and V-axis & mm/pixel \\ \hline
    $L_{sd}$ & Distance from source to detector & mm \\ \hline
    $L_{so}$ & Distance from source to object (Z-axis) & mm \\ \hline
    \end{tabular}
}
\label{tbl:parameters}
\end{table}

%% file: Algorithms/backprojection.tex
\SetInd{0.15em}{0.55em}
\begin{algorithm}[t]
\begin{minipage}{0.96\linewidth}
\caption{Back-projection algorithm for CBCT. \textit{interp2} function is used for interpolation~\cite{russ1990image}.}
\label{alg:BP}
\LinesNumbered
\KwIn{${\bf{Q}[P_n][P_r][P_c]}$ is filtered projection, ${\bf{M}[P_n]}$ is projection matrix (as details in ${M}_{\psi}$).}
\KwOut{${\bf{I}}[V_z][V_y][V_x]$ is output volume data.}
I $\gets$ 0 \Comment{initialize output, a 3D volume}

\For{each t $\gets$ 0 to $P_n$}{

    \For {each k $\gets$ 0 to $V_z$}{
        
        \For {each j $\gets$ 0 to $V_y$}{
        
            \For {each i $\gets$ 0 to  $V_x$}{

                $[x, y, z]^T \gets \textbf{M}[t]\cdot [i, j, k, 1]^T$ \Comment{Projection} 
                
                $[x, y] \gets [x, y] \cdot 1/z$

                $I[k][j][i]{\gets}I[k][j][i] + interp2(Q[t], x, y) / z^2$

            }
            
        }
        
    }
    
}

  \SetKwProg{Fn}{Function}{}{\KwRet}
  \Fn{interp2{($J,x,y$)}}
  {
        $[s_u,s_v]{\gets}[\floor{x},\floor{y}]$ \Comment{floor operation}\
        
        $[\varepsilon_u,\varepsilon_v]{\gets}[x-s_u, y-s_v]$\Comment{sub-pixel position}\
        
        $\alpha{\gets}J[s_v][s_u]{\cdot}(1-\varepsilon_u)+J[s_v][s_u+1]{\cdot}\varepsilon_u$\Comment{interp}\
        
        $\beta{\gets}J[s_v+1][s_u]{\cdot}(1-\varepsilon_u)+J[s_v+1][s_u+1]{\cdot}\varepsilon_u$\Comment{interp}\
        
        \KwRet $\alpha{\cdot}(1-\varepsilon_v)+\beta{\cdot}\varepsilon_v$\Comment{interp}\
  }
\end{minipage}
\end{algorithm}

%% file: Algorithms/Block.tex
\SetInd{0.15em}{0.55em}
\begin{algorithm}[t]
\small
\begin{minipage}{0.96\linewidth}
\caption{
BP algorithm integrates optimized memory-access patterns and Tensor Core acceleration. 
}
\label{alg:Block}
\LinesNumbered
\KwIn{$\mathbf{Q}[P_n][P_r][P_c]$, ${\bf{M}}[P_n][3][4]$, $\mathrm{L_{sd}}$ \Comment{as in Table~\ref{tbl:parameters}}} 
\KwOut{$\mathbf{I}[V_z][V_y][V_x]$. \Comment{as in Table~\ref{tbl:parameters}}}

block(tx, ty, 1)  \Comment{optimal thread Block in Sec.~\ref{sec:block}}

grid($\frac{V_x}{block.x}$, $\frac{V_y}{block.y}$, $\frac{V_z}{block.z}$) \Comment{optimal thread Grid in Sec.~\ref{sec:block}}

KernelLaunch$<<<$block, grid$>>>$(src, dst, M)

\_\_global\_\_ Function KernelLaunch(src, dst, M) :

\Indp
i\, = blockIdx.x $\cdot$ blockDim.x + threadIdx.x \Comment{i index in Alg.~\ref{alg:BP}}

j\, = blockIdx.y $\cdot$ blockDim.y + threadIdx.y \Comment{j index in Alg.~\ref{alg:BP}}

kk = blockIdx.z $\cdot$ blockDim.z $\cdot$ tz \Comment{base k index in Alg.~\ref{alg:BP}}







\For{each $t \gets 0$ to $P_n$ / 8 - 1 }{
    \For{each $k \gets kk$ to $kk + bz$}{
        
        $\mathbf{\_\_register}\ A[2][2][2], B[2][4][2], C[2][8][4]$ \
        
        
        \textbf{FP32toFP16}(M, $B_{high}$, $B_{low}$)  \Comment{\textcolor{blue}{as in Listing~\ref{listing:FP32toFP16}}} \
        
        $B[0], B[1] \gets B_{high}, B_{low} $   \
        
        \For{$m \gets 0$ \KwTo $1$}{
            \For{$n \gets 0$ \KwTo $3$}{

                \tcc{\textcolor{blue}{TC {MMA.m16n8k8} in Fig.~\ref{fig:fp16}} }
                
                $\mathrm{\bf MMA}(A[0][m][i], B[0][n][i], C[0][4 \cdot m + n][i])$ 
                
                $\mathrm{\bf MMA}(A[0][m][i], B[1][n][i], C[1][4 \cdot m + n][i])$ 
                
                $\mathrm{\bf MMA}(A[1][m][i], B[0][n][i], C[1][4 \cdot m + n][i])$ 
            }
        }
        
        
        \For{$p \gets 0$ \KwTo $7$}{
            $\hat{z} \gets C[i][4 \cdot \lfloor p/4 \rfloor][p \% 4]$ 
            
            $x \gets C[i][4 \cdot \lfloor p/4 \rfloor + 1][p \% 4] \cdot \hat{z} \cdot \mathrm{L_{sd}}$ \Comment{line 7 of Alg.~\ref{alg:BP}}\
            
            $y \gets C[i][4 \cdot \lfloor p/4 \rfloor + 2][p \% 4] \cdot \hat{z} \cdot \mathrm{L_{sd}}$ \Comment{line 7 of Alg.~\ref{alg:BP}}\
            
           $I[k][j][i]{\gets}I[k][j][i] + interp2(Q[t], x, y) / z^2$
        }
    }
}
\end{minipage}
\end{algorithm}

%% file: listing/decomposition.tex
\lstset{
 	language = C++, 
  basicstyle=\ttfamily\small, 
  keywordstyle = {\bfseries \color[cmyk]{0,1,0,0}}, 
  commentstyle = {\itshape \color[cmyk]{1,0.4,1,0}}, 
  stringstyle = {\ttfamily \color[rgb]{0,0,0}},
  numbers=left,
  numberstyle=\small,
  numbersep=3pt, 
  breaklines=true, 
  breakindent = 0pt, 
  lineskip={-0pt},
  columns=flexible, 
  keepspaces=true,
  breaklines=true,  
  xleftmargin=5pt,  
  xrightmargin=5pt,  
  showspaces=false,                
  showstringspaces=false,
  showtabs=false,    
  language=C++,               
  showstringspaces=false,
  showtabs=false,  
  frame=tb, 
  morekeywords={__volatile__, asm, mma, CVT_toFP16, \#define, __half, in, high, low},
  keywordstyle=\color{purple},
}
\lstset{escapeinside={<@}{@>}}

\begin{figure}[t]
\centering
\begin{minipage}[c]{0.5\textwidth}
\begin{lstlisting}[caption = {FP32-to-FP16 decomposition function in \method{}. This approach split a single FP32 value into two FP16 components to maintain the precision used in Alg.~\ref{alg:Block}.}, label = listing:FP32toFP16]
#define <@\textbf{FP32toFP16}@>(in, high, low)           \
  asm __volatile__ (                        \     
    ".reg.b16 h_val;" ".reg.b32 f_hi, f_lo;"\
    "cvt.rn.f16.f32 h_val, %2;"             \
    "cvt.f32.f16 f_hi, h_val;"              \
    "sub.f32 f_lo, %2, f_hi;"               \
    "mov.b32 %0, f_hi;" "mov.b32 %1, f_lo;" \
    : "=r"(high), "=r"(low) : "f"(in));

\end{lstlisting}
\end{minipage}
\end{figure}

%% file: tables/Dataset.tex
\setlength{\tabcolsep}{0.8pt}

\begin{table}[t]
    \centering
    \tiny
    \caption{Datasets used for evaluation.}
    \resizebox{\linewidth}{!}
    {
        \begin{tabular}{|c|c|c|c|c|c|c|c|c|c|c|c|}
        \hline
        \textbf{ Name} & \textbf{ $P_n$} & \textbf{ $P_c$} & \textbf{ $P_r$} & \textbf{ $\lambda_c(\lambda_r)$} & 
        \textbf{ $L_{sd}$} & \textbf{ $L_{so}$} & \textbf{ $\beta$} & \textbf{ $\omega_c$} & 
        \textbf{ $\omega_r$} & \textbf{ $\omega_{\text{cor}}$} \\ \hline
          Bumblebee &  3142 &  2000 &  2000 &  0.2 &  672.5 &  39.8 &  16.9 &  0 &  0 &  1.03 \\ \hline
      Tomo\_27 & \multirow{3}{*}{1800} & \multirow{3}{*}{1335} & \multirow{3}{*}{2004} & \multirow{3}{*}{0.025} & \multirow{3}{*}{250.0} & \multirow{3}{*}{100.0} & \multirow{3}{*}{2.5} & 25 & \multirow{2}{*}{0.25} & \multirow{3}{*}{0} \\ \cline{1-1} \cline{9-9}
        
         Tomo\_28 & & & & & & & & 26 &  & \\ \cline{1-1} \cline{9-10}
        Tomo\_29 & & & & & & & & 27 & 0.2 & \\ \hline
          Tomo\_30 &  720 &  445 &  668 &  0.075 &  350.0 &  250.0 &  1.4 &  -10 &  0.2 &  0 \\ \hline
        \end{tabular}
    }
    \label{tbl:datasets}
\end{table}     

%% file: tables/Performance.tex
\begin{table*}[t]
\scriptsize
\caption{End-to-end performance analysis of \method{} on Jetson Nano and Jetson AGX Orin.
$T_{Load}$ is the time required to load all projections; $T_{Filter}$ is the time for the filtering computation; $T_{BP}$ is the time for our optimized BP kernel; and $T_{Store}$ is the time to store the 3D volume.
The total time is $T_{sum}=T_{Load}+T_{Filter}+T_{BP}+T_{Store}$.
$T_{\method{}}$ denotes the end-to-end runtime of \method{} with overlapped I/O and computation, while $T_{RTK}$ denotes end-to-end runtime of the RTK library (baseline). 
$Speedup = T_{RTK}/T_{\method{}}$. 
“\xmark” indicates missing results due to out-of-memory in \textit{RTK}.
The representative examples of the pipeline are shown in Fig.~\ref{fig:jetson_nano_pipeline} and Fig.~\ref{fig:jetson_agx_pipeline}.
}
\centering
\resizebox{\linewidth}{!}
{       
    \setlength{\extrarowheight}{-15pt}
    \renewcommand{\arraystretch}{0.5}
    \begin{tabular}{|c|c|c|c|c||c|c|c|c|c||c|c||c|}
    \hline
    \rule{0pt}{6pt} 
    \textbf{Device} & \textbf{Dataset} & \textbf{Full Size} &  \textbf{Max Batch} & \textbf{Volume} & \textbf{$T_{load}$~(ms)} & \textbf{$T_{Filter}$~(ms)} &
    \textbf{$T_{BP}$~(ms)} & 
    \textbf{$T_{Store}$~(ms)} & \textbf{$T_{sum}$~(ms)} & \textbf{$T_{\method{}}$~(s)} & \textbf{$T_{RTK}$~(s)} & \textbf{Speedup~($\times$)}\\
    \hline
    \hline
    \multirow{26}{*}{\vspace{-9mm}\rotatebox[origin=c]{90}{\textbf{Jetson Nano}}}
    & \multirow{2}{*}[0ex]{Bumblebee} & \multirow{2}{*}[0ex]{47GB} & 2.98GB  &  $256^3$  & 1204469.1 & 10578.4 & 393.28 & 46.43 & 1215487.21 & 1531.99 & 2885.46 &  \textbf{1.88$\times$} \\
    &  &  & 3.15GB &  $512^3$& 1250684.2 & 10976.32 & 47647.11 & 31080.02 & 1340387.65 & 1621.48 & \xmark & \xmark \\
    \cline{2-13}
    & \multirow{4}{*}[-2ex]{Tomo\_29} & \multirow{4}{*}[-2ex]{17.9GB} & 0.25GB &  $256^3$  & 292194.12 & 8367.37 & 10141.27 & 141.53 & 310844.29 & 333.19 & 582.94 & \textbf{1.75$\times$} \\
    & &  & 0.25GB &  $512^3$ & 293682.69 & 8353.82 & 35391.39 & 825.33 & 338253.23 & 342.89 & 684.24 & \textbf{1.99$\times$} \\
    & &  & 2.00GB &  $1024^3$ & 291725.60 & 10021.49 & 196325.42 & 104217.90 & 602290.41 & 618.25 & \xmark & \xmark\\
    & &  & 2.00GB &  $2048^3$ & 1453768.77 & 41125.77  & 4590432.68 & 110274.88 & 6195602.09 &5203.03 &\xmark& \xmark\\
    \cline{2-13}
    & \multirow{4}{*}[-2ex]{Tomo\_30} & \multirow{4}{*}[-2ex]{816MB}  & 0.03GB &  $256^3$  & 1154.84 & 711.16 & 2675.71 & 153.99 &4695.70 & 4.72 & 7.02 & \textbf{1.49$\times$} \\
    & &  & 0.25GB &  $512^3$  & 1157.65 &  713.66 & 12481.94 & 854.93 & 15208.17 & 15.61 & 33.57 & \textbf{2.15$\times$} \\
    & &  & 0.33GB &  $1024^3$  & 15682.18 & 843.24 & 99328.36 & 86684.13 & 202537.91 & 208.78 & 370.11 & \textbf{1.77$\times$} \\
    & &  & 0.08GB &  $2048^3$  & 117846.09 &  2984.37 &3167547.96 & 139673.00 &3428051.41 &3388.19 &\xmark& \xmark\\
    \hline
    \hline
    \multirow{6}{*}{\vspace{-19mm}\rotatebox[origin=c]{90}{\textbf{Jetson AGX}}}
    & \multirow[c]{5}{*}[-3ex]{Tomo\_29} & \multirow{5}{*}[-3ex]{17.9GB}  & 0.25GB &  $256^3$  & 9288.98 & 3305.36  & 3831.78  &  73.51 & 16499.63 & 60.77 & 131.26 & \textbf{2.15$\times$} \\
    & &  & 0.25GB &  $512^3$ & 9363.25  & 2595.73 & 14025.70  & 508.79 & 26493.47 & 69.49 & 160.26 & \textbf{2.31$\times$}  \\
    & &  & 2.00GB &  $1024^3$ & 8716.80 & 2621.70 & 66576.77 &  4267.71 & 82182.37& 111.28 & \xmark & \xmark\\
    & &  & 2.00GB &  $2048^3$  & 97199.24  & 11153.79 &   502615.85 &  76528.15 & 687497.03 & 661.387  & \xmark & \xmark\\
    & &  & 2.00GB &  $3072^3$ &  226425.17  & 16587.02 & 1936609.82  & 182062.52 & 2361684.53 & 2084.15 & \xmark & \xmark\\
    \cline{2-13}
    & \multirow[c]{5}{*}[-3ex]{Tomo\_30} & \multirow{5}{*}[-3ex]{816MB}  & 0.03GB &  $256^3$ & 499.96 & 334.87 & 1023.20 &  83.76  & 1941.79 & 2.89 & 6.15 & \textbf{2.12$\times$} \\
    & &  & 0.25GB &  $512^3$ & 590.42 & 365.17 & 4403.05 & 553.95 & 5912.59 & 6.84 & 24.04 & \textbf{3.51$\times$} \\
    & &  & 0.33GB &  $1024^3$ & 4321.16 & 351.41 & 34727.58 & 4501.57 & 43901.72 & 44.24 & 121.79 & \textbf{2.75$\times$}  \\
    & &  & 0.33GB &  $2048^3$ & 57847.05 & 493.74 & 364523.91 & 50203.33 & 473068.03 & 430.57 & \xmark & \xmark  \\
    & &  & 0.25GB &  $3072^3$ & 178005.25 & 1348.30  & 1128073.21 & 155450.02  & 1462876.78 & 1104.09 & \xmark & \xmark\\
    \hline
    \end{tabular}
}
\label{tbl:performance}
\end{table*}

%% file: tex/2.bio.tex
\begin{IEEEbiography}[{\includegraphics[width=1in,height=1.25in,clip,keepaspectratio]{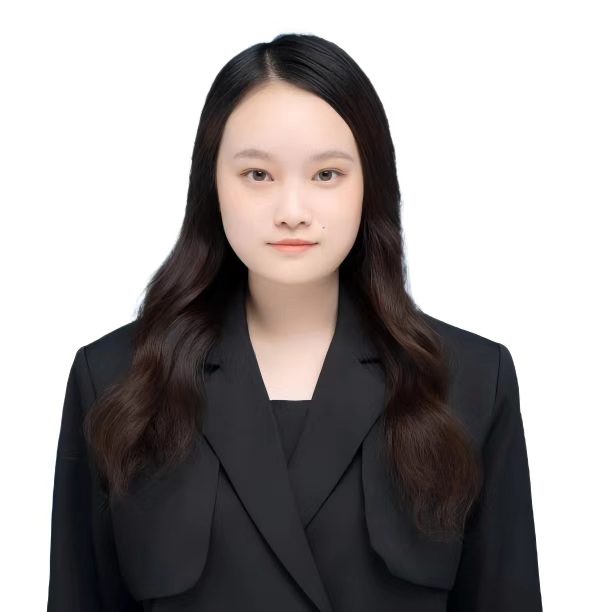}}]{Xuetao Chen}
Xuetao Chen is currently an M.Phil. student in the Department of Computer Science at Hong Kong Baptist University, Hong Kong. She received her B.Sc. degree in Software Engineering from Nankai University, Tianjin, China, in 2024. Her research interests include GPU programming, edge systems and heterogeneous runtime/scheduling.
\end{IEEEbiography}
\vspace{-30pt}

\begin{IEEEbiography}[{\includegraphics[width=1in,height=1.25in,clip,keepaspectratio]{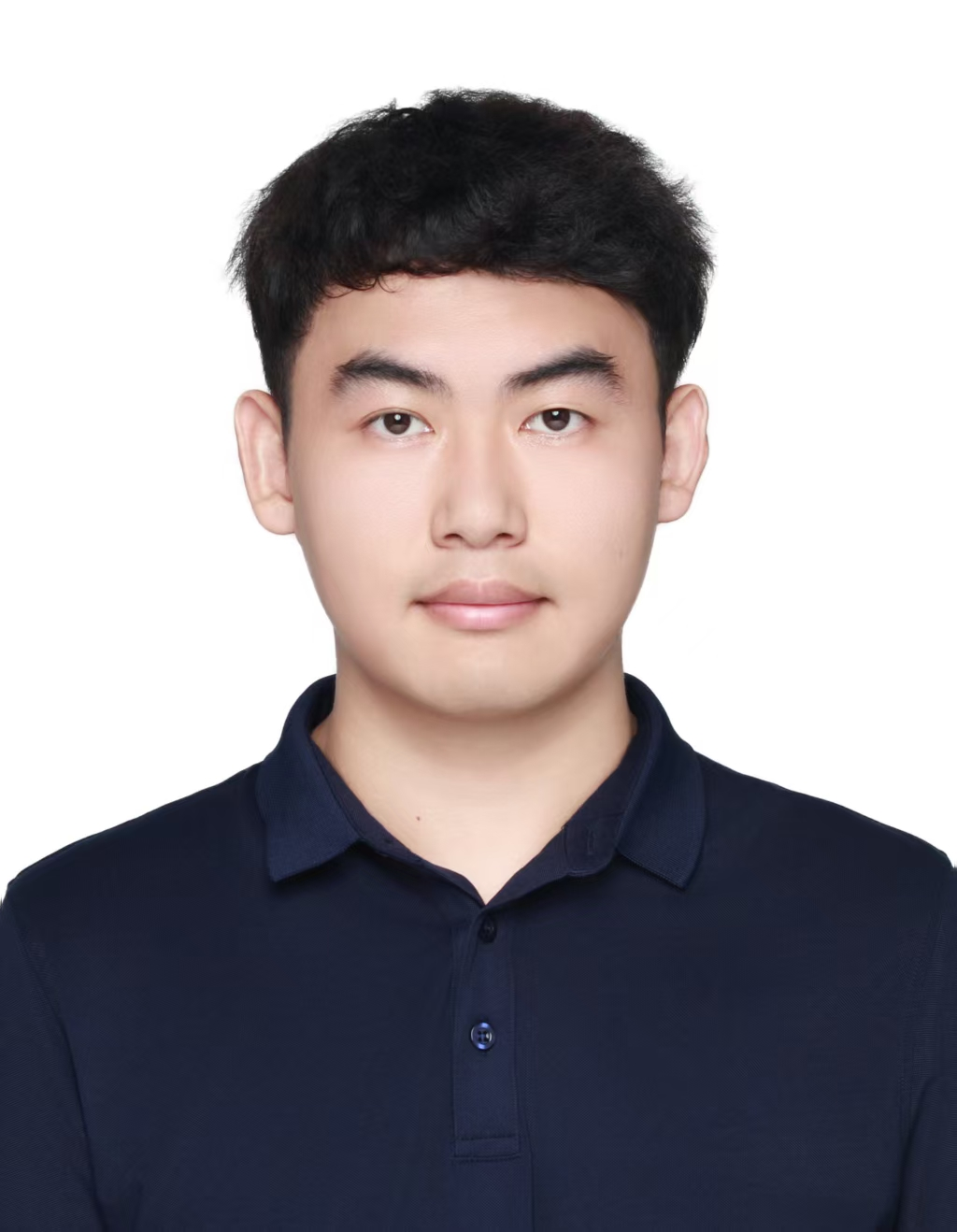}}]{Cong Ma}
Cong Ma is a Ph.D. student at the Graduate School of Information Science and Technology, Hokkaido University, Japan, and a Junior Research Associate at the RIKEN Center for Computational Science (RIKEN R-CCS), Japan. Prior to that, he received his M.E. degree in Computer Technology from the University of Chinese Academy of Sciences, China, in 2025, and his B.E. degree in Internet of Things Engineering from Southwest Petroleum University, China, in 2022. His research interests include high-performance computing, parallel computing, image processing.
\end{IEEEbiography}
\vspace{-30pt}

\begin{IEEEbiography}[{\includegraphics[width=1in,height=1.25in,clip,keepaspectratio]{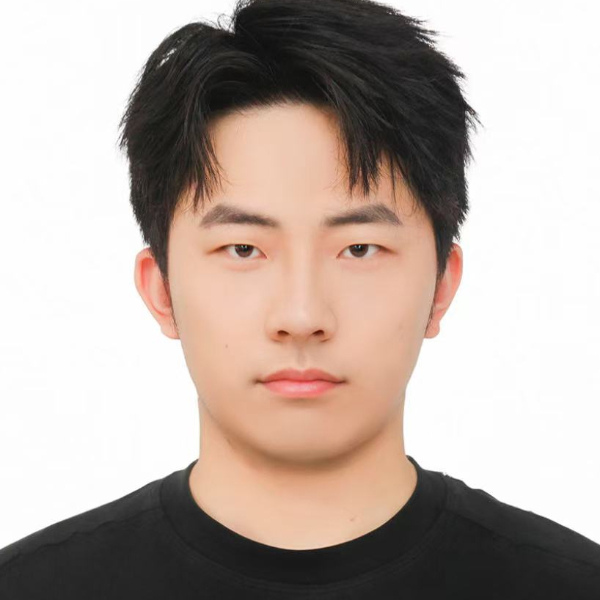}}]{Xiangyu Meng}
is a Ph.D. candidate at the College of Computer Science and Technology,  China University of Petroleum, Qingdao, China. His research interests include bioinformatics, parallel computing, and computational materials science.
\end{IEEEbiography}
\vspace{-30pt}

\begin{IEEEbiography}[{\includegraphics[width=1in,height=1.25in,clip,keepaspectratio]{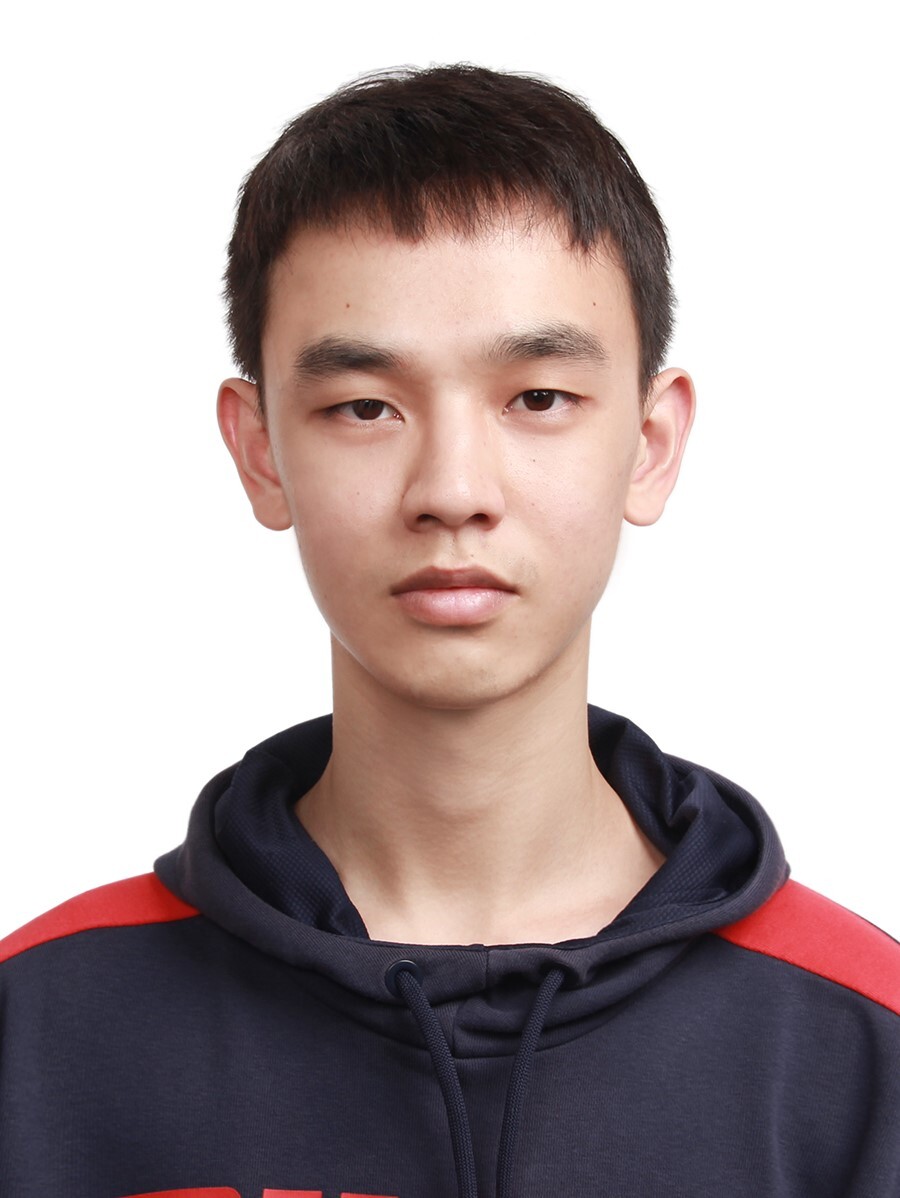}}]{Du Wu}
Du Wu is a Ph.D. student at the Institute of Science Tokyo (formerly Tokyo Institute of Technology), Japan, and a Junior Research Associate at the RIKEN Center for Computational Science (R-CCS), Japan. He received his B.E. degree in Computer Science and Technology from Jilin University, China, in 2020, followed by an M.E. degree in Electronic Science and Technology from the Southern University of Science and Technology, China, in 2023. His research interests include high-performance computing, performance portability across CPU and GPU architectures, and the optimization of dense and sparse matrix kernels for large-scale AI workloads. His work has appeared at top-tier venues such as SC, ICS, IPDPS, TPDS, and NeurIPS.
\end{IEEEbiography}

\vspace{-30pt}

\begin{IEEEbiography}[{\includegraphics[width=1in,height=1.25in,clip,keepaspectratio]{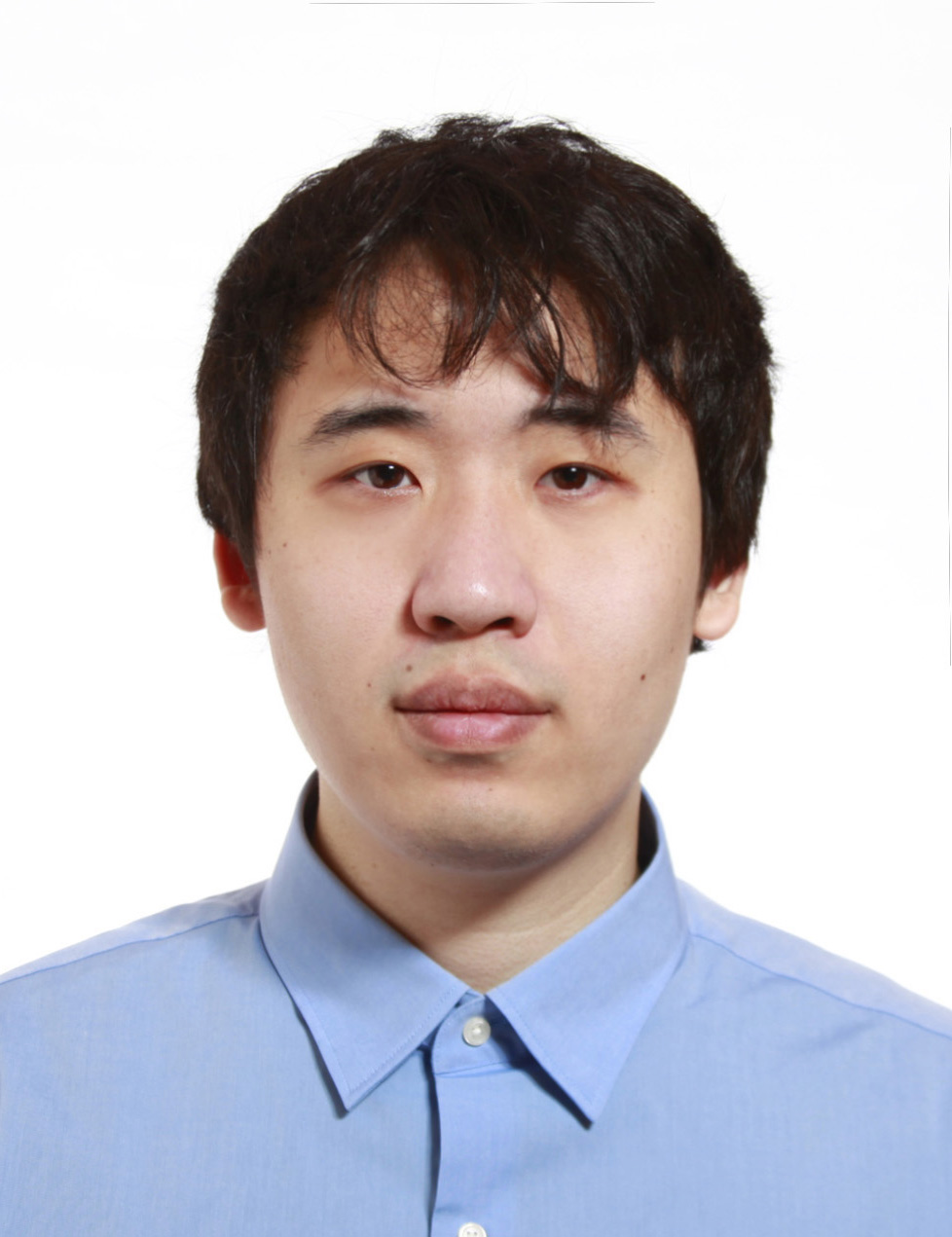}}]{Zhengyang Bai}
 received his B.E.\ in Software Engineering from East China Normal University in 2015,
and Master and Doctor degree of Informatics from Kyoto University in 2019 and 2023.
He is currently a postdoctoral researcher at the High Performance Artificial Intelligence System Research Team, RIKEN Center for Computational Science, Japan.
His research interests include high-performance computing, GPGPU and parallel programming languages. He is a member of IPSJ and ACM.
\end{IEEEbiography}

\vspace{-30pt}

\begin{IEEEbiography}[{\includegraphics[width=1in,height=1.25in,clip,keepaspectratio]{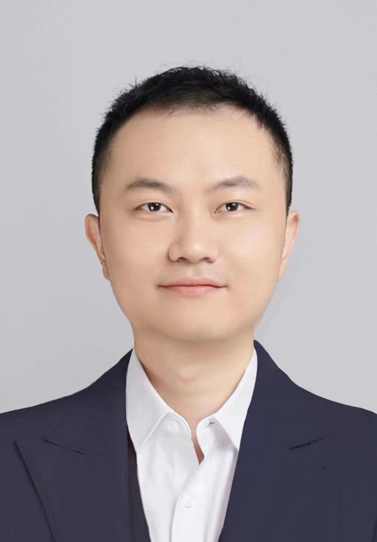}}]{Tao Luo}
Tao Luo (Senior Member, IEEE) received the B.S. degree from the Harbin Institute of Technology, Harbin, China, in 2010, the M.S. degree from the University of Electronic Science and Technology of China, Chengdu, China, in 2013, and the Ph.D. degree from the School of Computer Science and Engineering, Nanyang Technological University, Singapore, in 2018.
He is currently a Senior Research Scientist with the Institute of High Performance Computing, Agency for Science, Technology and Research (A*STAR), Singapore. He was an Associate Editor for the IEEE Transactions on Neural Networks and Learning Systems and is currently an Associate Editor for the IEEE Transactions on Computer-Aided Design of Integrated Circuits and Systems. His current research interests include high-performance computing, machine learning, computer architecture, hardware–software co-design, quantum computing, efficient AI, and their applications.
\end{IEEEbiography}

\vspace{-30pt}

\begin{IEEEbiography}[{\includegraphics[width=1in,height=1.25in,clip,keepaspectratio]{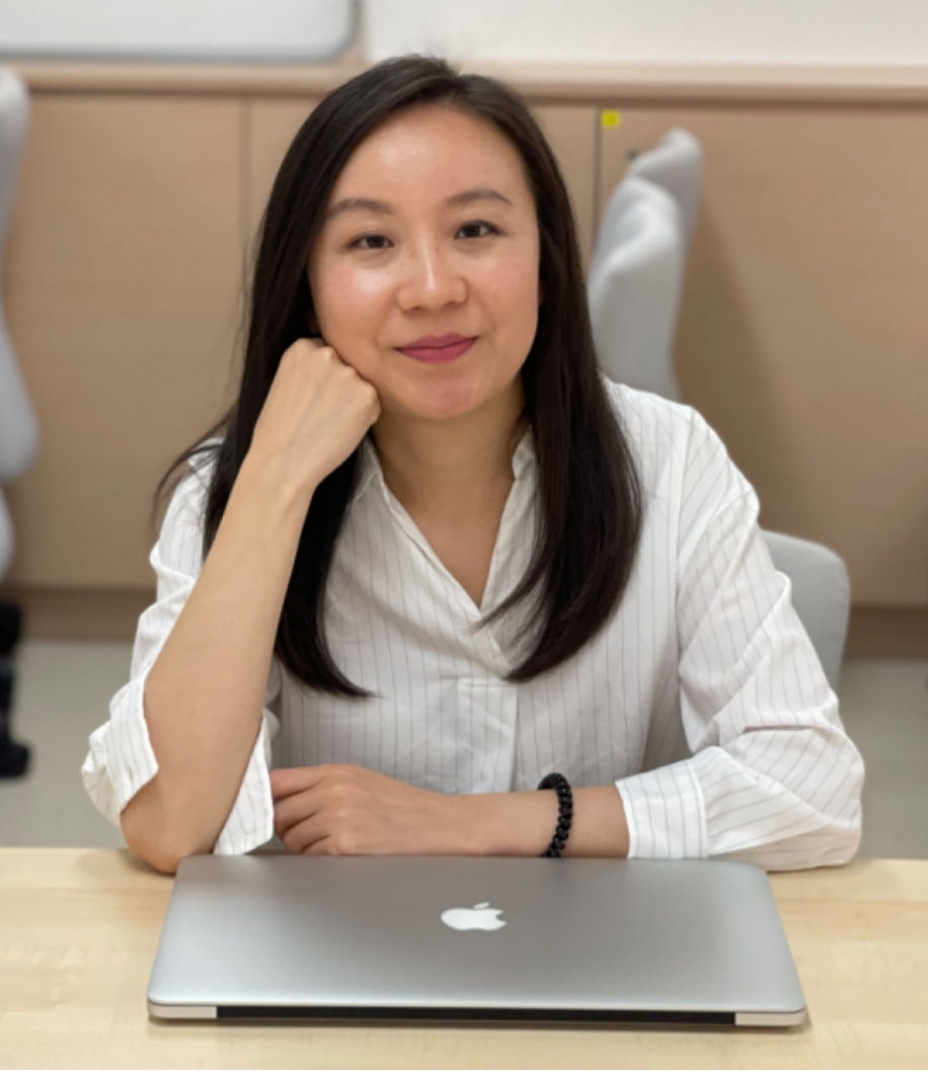}}]{Zhaorui Zhang}
Dr. Zhaorui Zhang is a research assistant professor at the Hong Kong Polytechnic University. She received her PhD degree from the University of Hong Kong and her bachelor's degree from Xi'an Jiaotong University. She worked in high-performance computing and AI infrastructure areas, specializing in system performance optimization across CPUs, GPUs, and FPGAs. Her recent research focuses on optimizing system performance for large-scale AI model training, fine-tuning, checkpointing, and inference, with particular emphasis on model compression and communication reduction. Dr. Zhang has published her work at numerous top-tier conferences and journals in the HPC field, including SC, IPDPS, ICCD, TPDS, IJCAI, and AAAI.
\end{IEEEbiography}

\vspace{-30pt}

\begin{IEEEbiography}[{\includegraphics[width=1in,height=1.25in,clip,keepaspectratio]{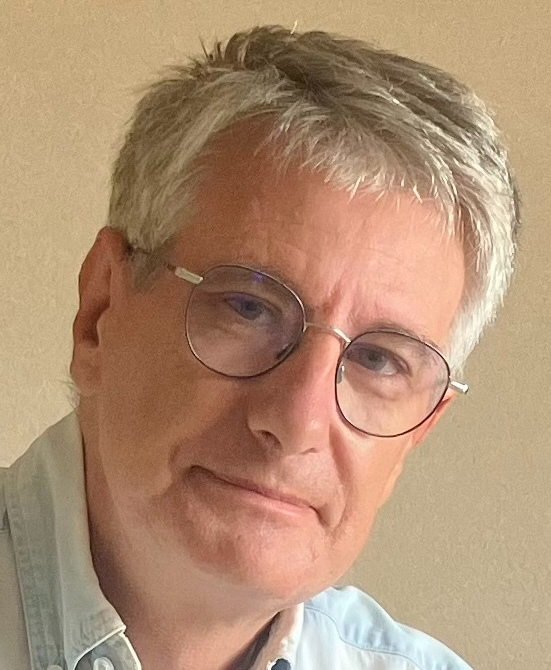}}]{Emmanuel Jeannot}
Emmanuel Jeannot is a Senior Research Scientist (Directeur de Recherche) at Inria, France.
He received his PhD in Computer Science from the École Normale Supérieure of Lyon in 1999 and his Habilitation (HDR) from Université Henri Poincaré, Nancy, in 2007.
His research spans over 25 years of contributions to high-performance computing, with a focus on parallel scheduling, topology-aware process placement, online compression, heterogeneous algorithms, and performance modeling for memory- and communication-bound workloads
From 2015 to 2024, Jeannot founded and led the TADaaM Inria research team (20 members).
Between 2024 and 2026, he joined DataDirect Networks Japan. During that period he was embedded within the HPAIS team at RIKEN R-CCS, where he developed GPU performance models for AI inference systems and designed a deadline-aware scheduler for large-scale model serving — broadening his expertise toward the intersection of HPC and AI infrastructure.
\end{IEEEbiography}

\vspace{-30pt}

\begin{IEEEbiography}[{\includegraphics[width=1in,height=1.25in,clip,keepaspectratio]{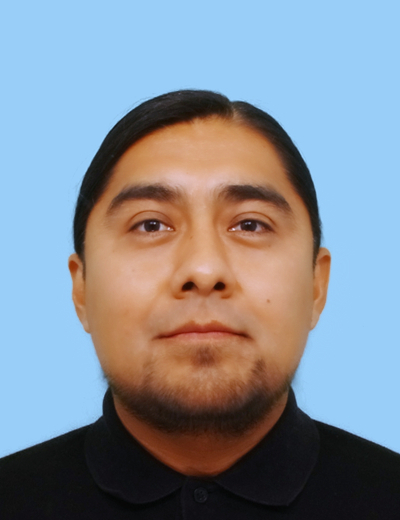}}]{Edgar Josafat Martinez-Noriega}
Edgar Josafat Martinez-Noriega obtained his Doctorate in Computer Science from the University of Electro-Communications, Tokyo in 2020. Following this, he has been employed as a Senior Researcher at the National Institute of Advanced Industrial Science and Technology (AIST), working on the application of synthetic datasets for large-scale deep learning. His research focuses on parallel computing, computer graphics, and deep learning.
\end{IEEEbiography}

\vspace{-30pt}

\begin{IEEEbiography}[{\includegraphics[width=1in,height=1.25in,clip,keepaspectratio]{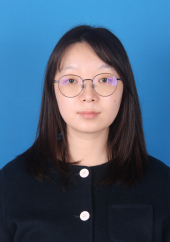}}]{Xun Wang}
Xun Wang received the Ph.D. from Tsukuba University. She is currently working as a Professor at the College of Computer Science and Technology of China University of Petroleum, Qingdao, China. Her research interests include bioinformatics, parallel computing, and computational materials science.
\end{IEEEbiography}

\vspace{-30pt}

\begin{IEEEbiography}[{\includegraphics[width=1in,height=1.25in,clip,keepaspectratio]{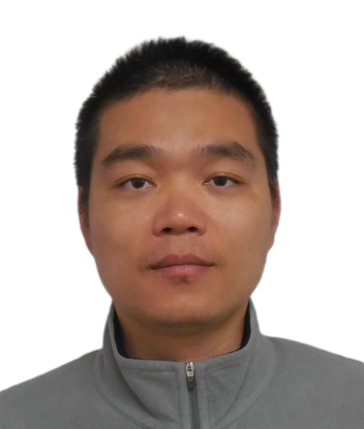}}]{Peng Chen}
Peng Chen is a senior scientist at the RIKEN Center for Computational Science (RIKEN-CCS), Japan. Prior to that, he was a researcher at the National Institute of Advanced Industrial Science and Technology (AIST), Japan. He received his B.E. degree in Navigation from Dalian Maritime University, China, in 2005, followed by an M.E. degree in Traffic Information Engineering and Control from Shanghai Maritime University, China, in 2007. He earned his Ph.D. from the Tokyo Institute of Technology, Japan, in 2020. His research interests include high-performance computing (HPC), parallel computing, image processing, evolutionary computation, and machine learning.
\end{IEEEbiography}

\vspace{-30pt}

\begin{IEEEbiography}[{\includegraphics[width=1in,height=1.25in,clip,keepaspectratio]{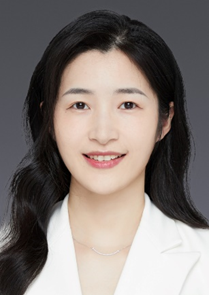}}]{Amelie Chi Zhou}
Amelie Chi Zhou received her PhD degree from Nanyang Technological University, Singapore, in 2016. She is currently an Assistant Professor with the Department of Computer Science, Hong Kong Baptist University (HKBU). Her research interests lie in high-performance computing, machine learning infrastructure, and memory-efficient computing architectures. She serves as an Associate Editor for IEEE TPDS, JPDC and FGCS. She is the recipient of the IEEE CS TCHPC Early Career Researchers Award.
\end{IEEEbiography}

\vspace{-30pt}


\begin{IEEEbiography}[{\includegraphics[width=1in,height=1.25in,clip,keepaspectratio]{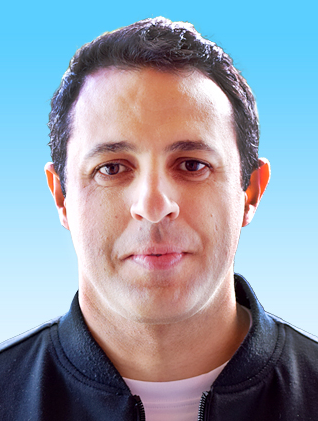}}]{Mohamed Wahib}
Mohamed Wahib is a team principal (PI) of the “High Performance Artificial Intelligence Systems Research Team” at RIKEN Center for Computational Science (R-CCS), Kobe, Japan. Prior to that he worked as a senior scientist at AIST/TokyoTech Open Innovation Laboratory, Tokyo, Japan. His research interests revolve around the central topic of high-performance programming systems, in the context of HPC and AI. He is actively working on several projects including AI-based science, as well as high-level frameworks for programming traditional scientific applications.
\end{IEEEbiography}
